\documentclass{aa}

\usepackage{graphicx}
\begin{document}  
\thesaurus{}
\title{New UBVRI colour distributions in E-type galaxies }
\subtitle{I.The data
\thanks{Based in part on observations collected at the Observatoire 
de Haute-Provence}}
\author{T.P. Idiart\inst{1},
  \and R. Michard\inst{2},
  \and J.A. de Freitas Pacheco\inst{3}}

\offprints{R. Michard} 
\institute{
Instituto de Astronomia, Geofisica e Ci\^encias Atmosf\'ericas, 
Depto. de Astronomia, 
Universidade de S\~ao Paulo, Av. Miguel Stefano, 4200-CEP 04301-904, 
S. Paulo, SP-Brazil; e-mail: thais@iagusp.usp.br
\and
Observatoire de Paris, LERMA, 77 av. Denfert-Rochereau,
F-75015, Paris, France; e-mail: Raymond.Michard@obspm.fr
\and
Observatoire de la Cote d'Azur, Dept. Augustin Fresnel
B.P. 4229, F-06304 Nice Cedex 4, France; e-mail: pacheco@obs-nice.fr}
\date{Received 17 September 2001 / Accepted 7 November 2001} 
\maketitle
\begin{abstract}
{New colour distributions have been derived from wide field UBVRI 
frames for 36 northern bright elliptical galaxies and a few lenticulars.   
The classical linear representations of colours against $\log r$ were derived, 
with some improvements in the accuracy of the zero point colours and of the 
gradients. The radial range of significant measurements was enlarged both 
towards the galaxy center and towards the outskirts of each object. Thus, 
the ''central colours", integrated within a radius of 3 \arcsec, and the 
''outermost colours" averaged near the $\mu_V=24$ surface brightness 
could also be obtained. 
Some typical deviations of colour profiles from linearity are described. 
Colour-colour relations of interest are presented. Very tight correlations are 
found between the U$-$V colour and the $Mg_2$ line-index, measured either at the 
galaxian center or at the effective radius.} 
\end{abstract}

\section{Introduction}

The ''classical" data on the large scale colour distributions of E-type 
galaxies relies on observations by Bender and M\"ollenhof (1987), 
Vigroux et al. (1988), 
Franx et al. (1989), Peletier et al. (1990), Goudfrooij et al. (1994), to 
quote only the papers discussing the 1-D profiles of colour against radius, as 
distinguished from studies of dust patterns. Most of these data were 
reconsidered by Michard (2000) (RM00), in an attempt to collect a significant 
sample of objects with a complete optical colour set, i.e. U$-$B, B$-$V, 
B$-$R and V$-$I in a coherent photometric system. This was adequate to 
confirm previous indications about the cause of colour gradients: these 
appear to be due essentially to population gradients within galaxies, with  
the dust playing no important role, except in galaxies with central intense 
dust patterns. Such objects are rather rare among the Es. 

Similar to most spectral indices of stellar populations, 
the colours suffer from the well known 
age-metallicity degeneracy, and, except U$-$B or U$-$V, are not very sensitive 
to the two parameters. They are affected by dust, at least locally, or perhaps 
systematically in the central regions according to inferences based 
on a survey by Michard (1999) (RM99). 
On the other hand, they may be measured at lower surface brightnesses or larger 
 radii than the line indices. They could therefore bring 
useful information to the study of fossil stellar populations, and further 
constraints upon models of the evolution of E galaxies. The present work 
aims to provide an enlarged sample of objects with complete colour 
data, extending farther in radius than in previous studies, and hopefully of 
improved accuracy. 

In Paper I, we present the usual information about the 
observations and data reduction, and part of the results in tabular form. 
A larger set of results will be made available in electronic form. The 
frames, partly reduced, will be made available from the HYPERCAT database,  
Observatoire de Lyon. 

In Paper II, under the assumption that the observed colour gradients 
reflect abundance variations along the radius, metallicity gradients will
be computed from the present data, using new colour-metallicity 
calibrations derived from multi-population models for E-galaxies. 
These metallicity gradients allow an estimation of central and mean
 metallicities. Statistics of galaxies included in our sample indicate
that mean metallicities are about solar, in agreement with the study
by Trager et al. (2000) based on spectral indices.  

\vspace{0.5 cm}

{\em Often used notations}
\begin{itemize}
\item $r$ isophotal radius; $r=(ab)^{1/2}$ for an ellipse of semi-axis $a$ and $b$.
\item $\Delta_{UB}$ colour gradient in U$-$B; 
$\Delta_{UB}=\mathrm{d(U-B)/d}(\log r)$ and similar for other colours.
\item diE, boE, unE: subclassification of E galaxies as disky, boxy and 
undetermined.
\end{itemize}

\section{Observations}

The observations were performed with the 120cm Newtonian telescope of the 
Observatoire de Haute-Provence, in three runs: April 1-11 2000, May 29-June 5 
2000 and January 18-29 2001, noted below as {\em run 1, 2 and 3}.  
Tables 1 to 4 gives lists of the observed galaxies 
with some parameters relevant on the observing conditions.
A CCD target Tek1024 is mounted in the camera, giving a field of view of 
11.6x11.6 \arcmin for a pixel size of 24 microns or 0.68 \arcsec.
The relatively large field is a favorable feature of this system for 
the observation of colour distributions in nearby galaxies. Less 
favourable are the rather poor seeing at the OHP, with the FWHM of star 
images usually in the 2-3\arcsec range, with values at 4 or more during 
periods of northern wind (mistral), and also the sky illumination by ever 
increasing urban lights. 

The camera is unfortunately affected by the so-called ''red-halo" effect. 

\begin{table}
\caption[ ]{A list of the observations. The dates refer to the begining 
of the night. Two values $W_o$ and $W_i$ are given for the frame FWHM 
(in \arcsec): the first is the original one; the second was attained 
after the treatment tending to equalize the FWHM of all 5 frames in 
a colour set, this at the value of the best one. Notes: 2974 mistral, interfering 
bright star; 3115 mistral; 3193 interfering bright star; 3605 mistral, 3607  
also observed; 3608 NGC3607 interfering. }
\begin{flushleft}
\begin{tabular}{llllllll}
\hline      

 NGC & Date & F & File & Exp & Sky & $W_o$ & $W_i$  \\
\hline
2768 & 01/04/00 & U & t613 & 3000 & 21.50 & 2.32 & 2.14  \\
id   & id       & B & t614 & 600  & 21.71 & 2.22 & 2.11  \\ 
id   & id       & V & t615 & 300  & 20.59 & 2.14 & -     \\
id   & id       & R & t616 & 240  & 19.93 & 2.29 & 2.14  \\
id   & id       & i & t617 & 240  & 18.86 & 2.36 & 2.15  \\
\hline
2974 & 05/04/00 & U & t828 & 3000 & 20.75 & 5.10 & 4.10  \\   
id   & id       & B & t830 & 600  & 21.47 & 4.63 & 4.05  \\
id   & id       & V & t829 & 300  & 20.54 & 4.04 &  -    \\
id   & id       & R & t831 & 240  & 20.11 & 4.58 &  -    \\
id   & id       & i & t832 & 240  & 18.47 & 4.23 &  -    \\
\hline
3115 & 20/01/01 & U & p437 & 2400 & 20.45 & 5.28 & 3.96  \\
id   & id       & B & p438 & 600  & 21.35 & 4.12 & 3.74  \\
id   & id       & V & p439 & 300  & 20.44 & 3.81 &  -    \\
id   & id       & R & p440 & 200  & 19.93 & 3.86 &  -    \\
id   & id       & i & p441 & 160  & 18.74 & 3.78 &  -    \\
\hline
3193 & 01/04/00 & U & t619 & 3000 & 21.09 & 3.11 & -  \\ 
id   & id       & B & t620 & 600  & 21.60 & 3.23 & -  \\
id   & id       & V & t621 & 300  & 20.78 & 3.26 & -  \\ 
id   & id       & R & t622 & 240  & 20.03 & 2.90 & -  \\
id   & id       & i & t623 & 240  & 19.27 & 2.75 & -  \\
\hline
3377 & 20/01/01 & U & p443 & 2300 & 21.10 & 3.64 & 2.81  \\
id   & id       & B & p444 & 600  & 21.72 & 3.02 & 2.69  \\
id   & id       & V & p445 & 300  & 20.72 & 2.73 &  -    \\
id   & id       & R & p446 & 200  & 20.16 & 2.80 &  -    \\
id   & id       & i & p447 & 160  & 18.85 & 2.76 &  -    \\
\hline
3377 & 27/01/01 & U & p674 & 2500 & 20.73 & 3.62 & 2.75  \\
id   & id       & B & p675 & 600  & 21.45 & 2.96 & 2.67  \\
id   & id       & V & p676 & 250  & 20.47 & 2.70 &  -    \\
id   & id       & R & p677 & 160  & 29.96 & 2.78 &  -    \\
id   & id       & i & p678 & 130  & 18.68 & 2.71 &  -    \\
\hline
3379 & 20/01/01 & U & p449 & 2500 & 21.28 & 3.96 & 2.90  \\
id   & id       & B & p450 & 600  & 21.86 & 3.29 & 2.84  \\
id   & id       & V & p451 & 250  & 20.89 & 2.91 &  -    \\
id   & id       & R & p452 & 160  & 20.32 & 3.11 &  -    \\
id   & id       & i & p453 & 130  & 19.07 & 2.99 &  -    \\
\hline
3605 & 05/04/00 & U & t842 & 3000 & 21.16 & 4.62 & 4.05  \\
id     & id       & B & t843 & 600  & 21.73 & 4.13 &  -  \\
id     & id       & V & t844 & 300  & 20.76 & 3.93 &  -   \\
id     & id       & R & t845 & 240  & 20.35 & 3.78 &  -   \\
id     & id       & i & t846 & 240  & 19.35 & 3.88 &  -   \\  
\hline
3608  & 06/04/00 & U & t937 & 3000 & 21.04 & 3.44 & 2.71  \\
id    & id       & B & t938 & 600  & 21.77 & 2.90 & 2.71  \\
id    & id       & V & t939 & 300  & 20.79 & 2.82 & 2.67  \\
id    & id       & R & t940 & 240  & 20.29 & 2.79 & 2.74  \\
id    & id       & i & t941 & 240  & 19.25 & 2.69 &  -   \\
\hline
3610 & 20/01/01 & U & p456 & 2500 & 21.39 & 2.99 & 2.49  \\
id   & id       & B & p457 & 600  & 21.99 & 3.11 & 2.51  \\
id   & id       & V & p458 & 250  & 21.07 & 2.71 & 2.44  \\
id   & id       & R & p459 & 160  & 20.49 & 2.51 &  -    \\
id   & id       & i & p460 & 130  & 19.21 & 2.75 & 2.43  \\
\hline
\end{tabular}
\end{flushleft}
\end{table}

\begin{table}
\caption[ ]{A list of the observations (continued). See conventions above. 
Notes: 4125 clouds; 4278 NGC4283 also observed; 4387 mistral; 4406 (30/5) 
clouds; 4406 (31/5) mistral.}
\begin{flushleft}
\begin{tabular}{llllllll}
\hline
3613 & 20/01/01 & U & p462 & 2500 & 21.25 & 3.58 & 2.43  \\
id   & id       & B & p463 & 600  & 21.92 & 3.16 & 2.25  \\
id   & id       & V & p464 & 250  & 20.90 & 2.42 & 2.11  \\
id   & id       & R & p465 & 160  & 20.26 & 2.16 &  -    \\
id   & id       & i & p466 & 130  & 19.03 & 2.20 &  -    \\
\hline
3640  & 07/04/00 & U & t040 & 3000 & 20.68 & 3.00 & 2.14  \\
id    & id       & B & t041 & 600  & 21.27 & 2.29 & 2.14  \\
id    & id       & V & t042 & 300  & 20.40 & 2.14 &  -    \\
id    & id       & R & t043 & 240  & 20.04 & 2.22 & 2.18  \\
id    & id       & i & t044 & 240  & 19.06 & 2.26 & 2.09  \\
\hline
3872  & 06/04/00 & U & t953 & 3000 & 21.03 & 2.89 & 2.36  \\
id    & id       & B & t954 & 600  & 21.68 & 2.49 & 2.43  \\
id    & id       & V & t955 & 300  & 20.70 & 2.38 &  -    \\
id    & id       & R & t956 & 240  & 20.17 & 2.49 & 2.37  \\
id    & id       & i & t957 & 240  & 19.18 & 2.67 & 2.43  \\
\hline
4125 & 21/01/01 & U & p521 & 2500 & 20.07 & 2.13 & 1.93   \\
id   & id       & B & p522 & 720  & 21.08 & 2.14 & 1.94   \\
id   & id       & V & p523 & 420  & 19.85 & 1.97 &  -     \\
id   & id       & R & p524 & 280  & 18.47 & 2.48 & 1.98   \\
id   & id       & i & p525 & 250  & 17.40 & 2.11 & 2.00   \\
\hline
4261 & 21/01/01 & U & p527 & 2500 & 20.05 & 2.84 & 2.39  \\
id   & id       & B & p528 & 720  & 21.10 & 2.61 & 2.40  \\
id   & id       & V & p529 & 420  & 20.25 & 2.41 &  -    \\
id   & id       & R & p530 & 280  & 20.07 & 2.61 & 2.50  \\
id   & id       & i & p531 & 250  & 18.78 & 2.53 & 2.52  \\
\hline
4278 & 25/01/01 & U & p600 & 2500 & 20.96 & 3.20 & 2.93  \\
id   & id       & B & p601 & 600  & 22.03 & 2.89 &  -    \\
id   & id       & V & p602 & 250  & 21.10 & 3.54 & 2.96  \\
id   & id       & R & p603 & 160  & 20.33 & 3.60 & 2.98  \\
id   & id       & i & p604 & 130  & 19.05 & 3.23 & 2.86  \\
\hline
4365 & 25/01/01 & U & p606 & 2500 & 20.72 & 3.52 & 2.93  \\
id   & id       & B & p607 & 600  & 21.75 & 3.39 & 2.95  \\
id   & id       & V & p608 & 250  & 20.74 & 2.99 &  -    \\
id   & id       & R & p609 & 160  & 20.01 & 3.17 & 2.94  \\
id   & id       & i & p610 & 130  & 18.64 & 2.82 &  -    \\
\hline
4374 & 01/04/00 & U & t629 & 3000 & 21.20 & 3.08 & -   \\ 
id   & id       & B & t630 & 600  & 21.84 & 3.20 & -   \\ 
id   & id       & V & t631 & 300  & 20.85 & 3.05 & -   \\
id   & id       & R & t632 & 240  & 20.36 & 2.94 & -   \\
id   & id       & i & t633 & 240  & 19.32 & 2.85 & -   \\
\hline
4387 & 05/04/00 & U & t856 & 3000 & 21.23 & 4.40 & -   \\
id   & id       & B & t857 & 600  & 21.79 & 4.35 & -   \\
id   & id       & V & t858 & 300  & 20.77 & 4.08 & -   \\
id   & id       & R & t859 & 240  & 20.35 & 4.23 & -   \\
id   & id       & i & t860 & 240  & 19.35 & 4.19 & -   \\  
\hline
4406 & 30/05/00 & U & m606 & 2400 & 20.06 & 2.65 & 2.14  \\
id   & id       & B & m607 & 600  & 20.69 & 2.69 & 2.19   \\
id   & id       & V & m608 & 300  & 19.21 & 2.65 & 2.12   \\
id   & id       & R & m609 & 150  & 17.76 & 2.39 & 2.09   \\
id   & id       & i & m610 & 120  & 17.28 & 2.09 & -     \\
\hline
4406 & 31/05/00 & U & m616 & 2400 & 20.42 & 4.66 & 2.14  \\
id   & id       & B & m617 & 500  & 20.50 & 4.27 & 2.19   \\
id   & id       & V & m618 & 250  & 19.24 & 4.32 & 2.12   \\
id   & id       & R & m619 & 180  & 19.08 & 4.14 & 2.09   \\
id   & id       & i & m620 & 150  & 18.16 & 3.85 & 2.09   \\
\hline
\end{tabular}
\end{flushleft}
\end{table}

\begin{table}
\caption[ ]{A list of the observations (continued). See conventions above.
Notes: 4551 NGC4550 also observed; 4552 clouds. }
\begin{flushleft}
\begin{tabular}{lllllllll}
\hline
4472  & 07/04/00 & U & t053 & 3000 & 20.98 & 2.43 & 2.29 \\
id    & id       & B & t054 & 600  & 21.53 & 2.49 & 2.30  \\
id    & id       & V & t055 & 300  & 20.60 & 2.24 &  -    \\
id    & id       & R & t056 & 120  & 20.16 & 2.37 &  -    \\
id    & id       & i & t057 & 120  & 19.09 & 2.32 &  -    \\
\hline
4473 & 29/05/00 & U & m571 & 2700 & 20.73 & 2.88 &  -    \\
id   & id       & B & m572 &  600 & 21.61 & 3.01 &  -    \\
id   & id       & V & m573 &  300 & 20.70 & 2.99 &  -    \\
id   & id       & R & m574 &  180 & 20.22 & 2.99 &  -    \\
id   & id       & i & m575 &  150 & 18.99 & 2.88 &  -    \\
\hline
4478 & 25/01/01 & U & p600 & 2500 & 21.00 & 3.56 & 2.58  \\
id   & id       & B & p601 & 600  & 21.59 & 3.64 & 2.65  \\
id   & id       & V & p602 & 250  & 20.62 & 3.12 & 2.48  \\
id   & id       & R & p603 & 160  & 19.92 & 3.11 & 2.54  \\
id   & id       & i & p604 & 130  & 18.62 & 2.52 & 2.86  \\
\hline
4486 & 27/01/01 & U & p680 & 2500 & 20.86 & 4.35 & 3.67  \\
id   & id       & B & p681 & 600  & 21.67 & 3.98 & 3.59  \\
id   & id       & V & p682 & 250  & 20.70 & 3.92 & 3.71  \\
id   & id       & R & p683 & 160  & 20.15 & 4.03 & 3.72  \\
id   & id       & i & p684 & 130  & 18.92 & 3.62 &  -    \\
\hline
4494 & 05/04/00 & U & t871 & 3000 & 21.18 & 3.62 & 3.30  \\
id   & id       & B & t872 & 600  & 21.66 & 3.31 &  -    \\
id   & id       & V & t873 & 300  & 20.56 & 3.47 &  -    \\
id   & id       & R & t874 & 240  & 20.04 & 3.47 &  -    \\
id   & id       & i & t875 & 240  & 18.92 & 3.26 &  -    \\
\hline
4551   & 01/06/00 & U & m647 & 2400 & 20.69 & 2.48 & 2.16  \\ 
id     & id       & B & m648 & 500  & 21.44 & 2.58 & 2.15   \\
id     & id       & V & m649 & 250  & 20.59 & 2.40 & 2.19   \\
id     & id       & R & m650 & 180  & 20.33 & 2.16 & -      \\
id     & id       & i & m651 & 150  & 19.43 & 2.28 & 2.13   \\
\hline
4552 & 02/06/00 & U & m693 & 2400 & 19.67 & 2.92 & 2.49   \\
id     & id     & B & m694 & 500  & 21.16 & 2.79 & 2.48     \\
id     & id     & V & m695 & 250  & 20.26 & 2.50 &  -       \\
id     & id     & R & m696 & 180  & 19.90 & 2.82 & 2.51     \\
id     & id     & i & m697 & 150  & 18.96 & 2.99 & 2.50     \\
\hline
4564 & 27/01/01 & U & p686 & 2500 & 20.86 & 3.92 & 3.41  \\
id   & id       & B & p687 & 600  & 21.59 & 3.44 &  -    \\
id   & id       & V & p688 & 250  & 20.70 & 3.56 &  -    \\
id   & id       & R & p689 & 160  & 20.00 & 3.85 & 3.47  \\
id   & id       & i & p690 & 130  & 18.62 & 3.53 &  -    \\
\hline
4621 & 06/04/00 & U & t966 & 3000 & 20.75 & 2.85 &  -    \\
id   & id       & B & t967 & 600  & 21.50 & 2.81 &  -    \\
id   & id       & V & t968 & 300  & 20.39 & 3.05 & 2.77  \\
id   & id       & R & t971 & 120  & 19.68 & 2.82 &  -    \\
id   & id       & i & t972 & 120  & 19.24 & 2.88 &  -    \\
\hline
4636 & 27/01/01 & U & p693 & 2263 & 20.68 & 3.86 & 3.11  \\
id   & id       & B & p694 & 662  & 20.78 & 3.61 & 3.06  \\
id   & id       & V & p695 & 300  & 20.25 & 3.35 & 2.99  \\
id   & id       & R & p696 & 210  & 19.47 & 3.47 & 2.99  \\
id   & id       & i & p697 & 180  & 18.22 & 3.05 &  -    \\
\hline
4649 & 03/06/00 & U & m723 & 2400 & 20.15 & 2.57 &  -     \\
id   & id       & B & m724 & 500  & 21.20 & 2.62 &  -     \\
id   & id       & V & m725 & 250  & 20.18 & 2.52 &  -     \\
id   & id       & R & m726 & 180  & 19.82 & 2.52 &  -     \\
id   & id       & i & m727 & 150  & 18.80 & 2.34 &  -     \\
\hline
\end{tabular}
\end{flushleft}
\end{table}

\begin{table}
\caption[ ]{A list of the observations (continued). See conventions above. 
Notes: 5831 mistral and clouds.}
\begin{flushleft}
\begin{tabular}{lllllllll}
\hline
5322 & 06/04/00 & U & t992 & 3000 & 20.95 & 2.91 & 2.52   \\
id   & id       & B & t993 & 600  & 21.72 & 2.64 & 2.64   \\
id   & 07/04/00 & V & t994 & 300  & 20.69 & 2.93 & 2.53   \\
id   & id       & R & t080 & 240  & 20.62 & 1.80 & 2.51   \\
id   & id       & i & t081 & 240  & 19.49 & 1.87 & 2.51   \\
\hline
5576  & 07/04/00 & U & t083 & 3000 & 20.73 & 2.31 &  -    \\
id    & id       & B & t084 & 600  & 21.22 & 2.52 & 2.31  \\
id    & id       & V & t085 & 300  & 20.38 & 2.67 & 2.30  \\
id    & id       & R & t086 & 240  & 19.87 & 2.48 & 2.26  \\
id    & id       & i & t087 & 240  & 18.79 & 2.39 &  -    \\
\hline
5813 & 29/05/00 & U & m579 & 2400 & 20.95 & 3.47 & 3.18   \\
id   & id       & B & m580 & 500  & 21.66 & 3.28 & 3.18   \\
id   & id       & V & m581 & 250  & 20.71 & 3.19 & -      \\ 
id   & id       & R & m582 & 180  & 20.28 & 3.09 & -      \\
id   & id       & i & m583 & 150  & 19.13 & 3.16 & -      \\
\hline
5831 & 31/05/00 & U & m622 & 2400 & 20.43 & 3.89 & 3.28  \\
id   & id       & B & m623 & 500  & 20.66 & 4.07 & 3.28   \\
id   & id       & V & m624 & 250  & 18.84 & 3.33 &  -     \\
id   & id       & R & m625 & 180  & 18.06 & 3.23 &  -     \\
id   & id       & i & m626 & 150  & 16.73 & 3.19 &  -     \\ 
\hline
5846 & 01/06/00 & U & m654 & 2400 & 20.66 & 2.55 & 2.11   \\
id   & id       & B & m655 & 500  & 21.28 & 2.74 & 2.16   \\
id   & id       & V & m656 & 250  & 20.38 & 2.07 &  -     \\
id   & id       & R & m657 & 180  & 20.08 & 2.28 & 2.11   \\
id   & id       & i & m658 & 150  & 19.08 & 2.33 & 2.11   \\
\hline
5866 & 02/06/00 & U & m700 & 2400 & 21.16 & 2.43 & 1.93   \\
id   & id       & B & m701 & 500  & 21.76 & 2.26 & 1.88  \\
id   & id       & V & m702 & 250  & 20.80 & 1.84 &   -    \\
id   & id       & R & m703 & 180  & 20.36 & 1.91 &   -    \\
id   & id       & i & m704 & 150  & 19.37 & 1.82 &   -    \\
\hline
5982 & 03/06/00 & U & m730 & 2400 & 20.89 & 2.20 & 2.00  \\
id   & id       & B & m731 &  500 & 21.15 & 2.30 & 2.06  \\
id   & id       & V & m732 &  250 & 20.46 & 2.07 &  -    \\
id   & id       & R & m733 &  180 & 20.18 & 2.11 &  -    \\
id   & id       & i & m734 &  150 & 19.13 & 2.33 & 2.06  \\
\hline
\end{tabular}
\end{flushleft}
\end{table}
    
\section{Data analysis}
\subsection{Outline of the operations}
The analysis of the frames entails the following steps:
\begin{enumerate}
\item The usual corrections for offset, the ''flat-fielding", and the 
interpolation of bad columns. As explained below we tried to improve 
the flat-fields by ad hoc ''superflats".
\item The registration of the 5 frames in each passband to a common geometry, 
based on a set of measured coordinates for 6 to 12 stars. This is intended 
to simplify the derivation of colour maps if needed.
\item The preparation of each frame for measurement, involving a final attempt 
to measure and correct residual large scale background trends, corrections for
parasitic objects, a treatment against cosmic rays peaks, and the calibration 
against the available results of aperture photometry (see Sect. 3.2.3).
\item Large errors in colour measurements may result from small differences 
between the widths of the PSFs of the two frames involved (see for instance 
Michard 1999, and the previous literature quoted therein). When the FWHMs 
of measured PSFs in the colour set for a given object differed by more than 
10\%, it was our practice to modify the frame PSFs and try to make them equal to 
that of the best frame of the group (see Sect. 3.2.4). 
\item The isophotal analysis of the V frame was performed according to Carter   
(1978), as implemented in the Nice technique described in Michard and Marchal 
(1994) (MM94). 
\item The correction procedures for the red halo effect in V$-$I, and eventually 
for the effects of different PSF far wings in other colours, were performed
(see Sect. 3.2.5). 
The correction necessitates the crossed correlation of the V frame by the I PSF 
and conversely. This operation cancels out the errors in the colour distribution 
induced by the red halo, but degrades the resolution.  
A correction to the calibrations performed before the convolutions 
is needed. 
\item Finally, colour measurements were performed along the previously found 
isophotal contours, and the average isophotal colours tabulated. Our routine 
at this stage involves corrections to the adopted values of the sky backgrounds,  
in order to eliminate the obvious effects of inaccuracies in these 
(see Sect. 3.2.6). 
\item The V surface brightness and colours have been collected in ad hoc files, 
and corrected for galactic absorption (or reddening) and the K effect, according 
to the precepts and constants given in the Third Reference Catalogue of Bright 
Galaxies (RC3, de Vaucouleurs et al., 1991). The usual linear representation of 
colours against $\log r$ have been calculated {\em in selected ranges}, 
avoiding on the one hand the central regions affected by imperfect seeing 
corrections and/or by important dust patterns; and on the other the outer range 
presumably affected by poor background corrections and residual noise.  
                   
\end{enumerate}
\subsection{Detail of operations.}
Some important details of the above summarized procedures will now be discussed.

\subsubsection{The background} 
The sky background of flat-fielded frames showed disappointingly {\em 
large scale trends}, specially in the U band. 
To improve upon this situation, it was tried 
to derive corrections by mapping the background of such frames, at least those 
which were not ''filled" by a large galaxy. 
Such maps were found to be correlated, although less so 
than expected, and their average used as a ''superflat". The quality of the 
background was then generally improved: if not, the superflats were not used.  
Note that, in the observing runs of May 2000 
and January 2001, a number of frames of ''empty" fields were obtained (sometimes 
in moonlight hours) to contribute to the derivation of superflats. 

A final improvement was obtained by measuring, during the treatment  
of each frame, a number of background patches, and subtracting a linearly 
interpolated map of these, instead of a constant. 
The background residual {\em large scale} fluctuations 
were often measured at the three steps of the procedure, that is, after the 
application of the flats, of the superflats, and after the final treatment. 
Table 5 summarizes the results. It may be noted that the combination of flats 
and superflats left large errors in our first and second runs, specially in the 
U colour. The final background linear ''rectification" allowed quite significant 
improvements, as seen by comparing the upper and lower halves of the 
table. 
 
\begin{table}
\caption[ ]{Measurements of residual fluctuations in background before and 
after the final ''rectification". Unit: \% of sky background.}
\begin{flushleft}
\begin{tabular}{llllll}
\hline      
								
Colour                   &  U   &  B   &  V   &  R   &  i  \\	
\hline
Flats+superflats 1st run & 1.84 & 1.19 & 0.72 & 0.68 & 0.95 \\
id.              2nd run & 1.90 & 0.77 & 0.45 & 0.54 & 0.79 \\
id.              3rd run & 0.69 & 0.79 & 0.56 & 0.60 & 0.64 \\
\hline
Final treatment  1st run & 0.36 & 0.34 & 0.30 & 0.29 & 0.28 \\
id.              2nd run & 0.40 & 0.27 & 0.26 & 0.31 & 0.30 \\
id.              3rd run & 0.28 & 0.28 & 0.24 & 0.20 & 0.32 \\
\hline
\end{tabular}
\end{flushleft}
\end{table}

If we consider an E galaxy observed under 
the typical conditions of the present series (see above for a tabulation of 
sky background values), the final residual fluctuations quoted here 
represent {\em local errors} of less than 0.1 mag near the isophote $\mu_B=25$.  
We will return later to the question of errors resulting from background 
uncertainties. 

\subsubsection{Parasitic objects} In galaxy photometry it is necessary to  
remove parasitic objects, stars and galaxies, that overlap the measured 
object. In the present work we used concurrently the following techniques: 

	.  replacement of pixels in a circle enclosing the ''parasite" by a 
circle symmetric about the center of the galaxy.

	. replacement of the circle by another one chosen in a nearby area.

	. marking of the pixels to be discarded in such a way that they are 
later left aside in the measuring programmes.  

\subsubsection {Calibrations} The frames were calibrated by comparisons with 
the results of aperture 
photometry. Our first choice was to use the UBVRI data of Poulain (1988) (PP88), 
and Poulain and Nieto (1994) (PN94) that are available for 26 objects of our 
survey, and are in Cousins's system, notably for R and I.
In a few cases the data collected in the HYPERCAT catalogues were used. 
These contain both primary data (those with an independent photometric calibration), 
and secondary data (actually calibrated with part of the primary data). 
Only primary data were used, selected according to our previous experience or 
prejudices. A few completely missing calibrations in R and I were replaced 
by values calculated from the tight correlations of V$-$R and V$-$I with B$-$V, 
derived from Poulain's data for E galaxies.

Although the I filter in the camera is of Gunn's type, our photometry 
is transfered to Cousins's system through the calibration. It is assumed that the 
difference of pass-bands has no significant effect on colour gradients.

\subsubsection{Equalization of PSF FWHM's}
Many studies of colour distribution in galaxies are affected by errors resulting 
from the difference in the PSFs of the two frames involved in a colour measurement. 
Franx et al. (1989), Peletier et al. (1990), Goudfrooij et al. (1994) calculated
the radial range where the errors due to ''differential seeing" are larger than 
some accepted threshold, and discarded the corresponding colours. Our policy, for 
instance in RM99, was to correct for this effect by adjusting the two 
frames to have PSFs with a common FWHM.
 
In the present study we tried to equalize the PSFs of the 5 frames 
in a given colour set. This is feasible if the 5 frames are taken in rapid 
succession, 
so that the PSFs have similar widths. The frame with the narrowest PSF is 
selected, 
and we find by trial and error a narrow gaussian (or sum of two gaussians) which 
can restore another frame to the same quality, or rather the same FWHM, by 
deconvolution. 
The parameters at hand are the $\sigma$ of the gaussian and the number of 
iterations in the deconvolution. In Table 1 to 4 we give the 
original FWHM $W_o$ of each frame, and the improved $W_i$ after the  
procedure described here. 
Obviously this cannot lead to perfect results, and sometimes we find in our data 
the signature of ''differential seeing", in the form of {\em large} colour 
variations, peaks or dips, within the seeing disk: these defects were edited out 
unless there was some good reason to suspect a genuine central colour anomaly, 
such as large dust patterns, or the jet of NGC4486.  

The reader may notice that two observations of NGC4406 are listed 
in Table 2, 
one of May 30, the other of May 31 2000. The first one was taken through 
fog and with average seeing, while for the other the ''mistral" 
brought a clearer sky and very poor seeing. A special treatment was then 
applied: the central peak of the sharp images was ''grafted" on the 
corresponding regions of the unsharp but deeper images. This explains 
why the $W_i$ is so much narrower than the original $W_o$ for the frames 
of May 31.

\begin{table*}
\caption[ ]{Comparisons of average color gradients for different subsamples of 
E galaxies, before or after tentative corrections for the effects of PSF far 
wings according to RM01. No corrections are applied if the average observed 
gradients are close to the adopted reference.  
N, number of objects. $\Delta_{U-B}$, etc. mean gradients.
}
\begin{flushleft}
\begin{tabular}{llllllllll}
\hline      

 Subsample    & N & $\Delta_{U-B}$ & $\sigma$ & $\Delta_{B-V}$ & $\sigma$ &  
 $\Delta_{V-R}$ & $\sigma$ & $\Delta_{V-I}$ & $\sigma$ \\   
\hline
 RM00    & 29 & -0.152 & 0.048 & -0.061 & 0.025 & -0.018 & 0.030 & -0.053 
& 0.022  \\

 2000 Observ. & 23 & -0.138 & 0.037 & -0.064 & 0.018 & +0.018 & 0.015 & +0.093     
& 0.047  \\
2000 Correc.  & 23 & -0.116 & 0.038 &  -     & -     & -0.016 & 0.013 & -0.048 
& 0.026 \\

2001 Observ.  & 14 & -0.174 & 0.045 & -0.080 & 0.022 & -0.017 & 0.013 & +0.040  
& 0.037 \\
2001 Correc.  & 14 & -0.140 & 0.036 &  -     & -     &  -     & -     & -0.062
& 0.025  \\

\hline
\end{tabular}
\end{flushleft}
\end{table*}
\begin{figure}
  \resizebox{\hsize}{!}{\includegraphics{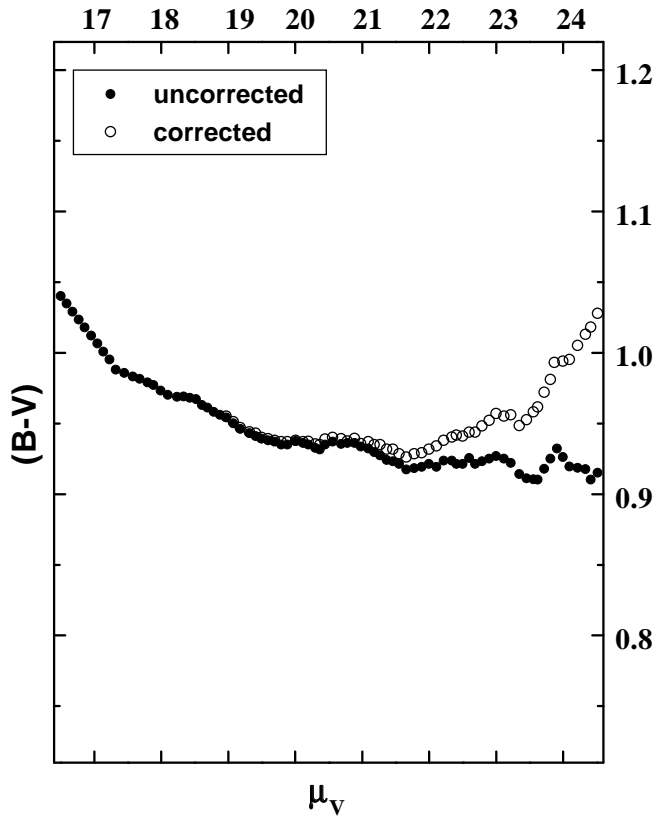}}
 \caption{Example of the ''correction" of a colour profile through changes 
in the sky background constants. Abscissae: V surface brightness in mag. 
Ordinates: B$-$V. Uncorrected: open circles. Corrected: filled circles. 
The changes of the background amount here to -0.25\% in B and 0.10\% in V, 
that is more than average (see Table 7)}
 \label{ }
\end{figure}
  
\subsubsection{ ''Red halo" and PSFs far wings.}
Before the start of this survey, the CCD camera on the telescope used was 
known to be affected by the ''red halo", an unfortunate property of thinned CCDs. 
The aureoles surrounding stellar images are obviously brighter 
and more extended in the I band than in B or V. Not only the red halo, but more 
generally the outermost wings of PSFs, were measured during our observing runs 
in 2000-1. The techniques and results are described in Michard (2001) (RM01). 
{\em The choice of appropriate star fields allowed us to extend the 
measurements up to a radius of nearly 3 \arcmin, and down to a level of about 
$0.5.10^{-6}$ of the central peak}. Due to the red halo effect, PSF wings in I may be a 
factor of 3 brighter in an extended radius range than the V ones. Much smaller 
but still significant differences may also occur between the PSFs of various 
spectral bands, the V PSF wings always being fainter. The V PSF wings however, 
and all the others at the same time, were reinforced between our observing runs 
of spring 2000 and winter 2001, probably an effect of 10 months ageing of mirrors 
coatings. The final output of the measurements are average ''synthetic" 
PSFs in the format 512x512 pixels, or 5.8x5.8 \arcmin, for each run and pass-band.
  
A set of numerical experiments on model galaxies is also presented 
in RM01, to illustrate the effects of these far wings on the 
observed surface brightness and colour distributions. The most striking effect 
occurs for the gradient $\Delta_{VI}$, which appears strongly positive, 
while it is negative according to the classical results of Bender and M\"ollenhof 
(1987), or Goudfrooij et al. (1994). More subtle effects are found for other 
colours, with relatively small but definite changes in gradients. 

To correct for the consequences of the red halo, or other similar effects 
upon the colour distribution in the index C1$-$C2, {\em frame C1 is convolved with 
the PSF of frame C2 and conversely.} 
After this operation, the resulting images have been submitted to the same 
set of convolutions, one in the atmosphere plus instrument, the other 
in the computer: they lead therefore to correct colour distributions, but 
with a significant loss of resolution. 
As the convolutions attenuate the central regions of the galaxy, and much 
more so for the V frame convolved by the I PSF, the mean colours are biased: 
a correction to the calibrations performed before the convolutions 
is needed. This has been done by a comparison of simulated aperture photometry 
to the observed one, an operation also used to estimate the errors in calibration 
(see below). 

Since the extended PSFs are found with limited accuracy, 
it is necessary to discuss the validity of the corresponding corrections 
obtained through crossed convolutions, the more so because of the obvious 
changes of the PSF far wings between {\em run 1} and {\em run 3}. 
The mean values of the colour gradients for subsamples of E galaxies have been 
used for these checks, with the results of Table 6. For a 
subsample of 12 or more E galaxies the mean colour gradients and their 
dispersions cannot differ much, so that their values may be used as checks of 
the need for a correction and its eventual success. 
\footnote {This assertion is questionable as pointed out by the referee.  
We therefore tested it by comparing the distributions of the gradients 
in subsamples of 15 to 25 objects, 
sorted out by NGC numbers, from the surveys of Peletier et al. 
1990, Goudfrooij et al. 1994 and our discussion in RM00. This procures 8 tests, 
according to the number of colours in each of the sources. It turns out that 
the subsamples are statistically coincident in 4 cases. They differ by  
slightly more than the calculated errors in the others.}  The reference for these 
comparisons is the subsample in Michard (2000) (RM00), mostly a rediscussion 
of the ''classical" data by Peletier et al. (1990), Goudfrooij et al. (1994) 
and others.

Looking at the Table 6, it is clear that the red halo introduces enormous errors 
in the V$-$I gradients, but that the corrections are remarkably successful 
in restoring the agreement of the results with the accepted reference, both as 
regards the mean values and the dispersion. The same may be said about the 
V$-$R gradients. The wings of the V PSF were strongly 
reinforced between our {\em run 3} and {\em run 1 or 2}, but much less so 
for the I and R PSF wings. As a result the red halo effect is less in 
V$-$I for the frames of {\em run 3} and disappears in V$-$R.   
  
The situation is less clear for the U$-$B distributions. Our mean uncorrected 
gradients are in good agreement with the ''classical" data, essentially from
Peletier et al. (1990), as rediscussed in RM00. 
On the other hand, the U PSF 
wings are consistently above the V ones in all our runs, so that the true 
slopes of the U$-$B variations may be a bit smaller than the observed ones. 
This error might well be present in the classical observations. 
Incidentally, the data of Peletier et al. were obtained in U$-$R, and it is 
impossible to be sure that the far PSF wings of the used telescope were the 
same in both pass-bands! 

Similar remarks might be made about our B$-$V data. Since the mean 
measured gradient is the same for our data of the year 2000 and the adopted 
reference (and as a good B PSF is not available) we take as correct this 
set of results. For our data of 2001, there is evidence that the wings 
of the B PSF were slightly above those of the V one, so that the B$-$V 
gradients might also be biased upwards. 

{\em It appears that a significant source of error in the measurement of the 
small colour gradients in E galaxies has hitherto been overlooked. 
It might be that small systematic errors, of the order of some 15-20\%, 
are still present in the U$-$B or U$-$V gradients published here.
Although such errors would not have significant astrophysical implications, 
control observations are planned.}  

\subsubsection{Measurements of isophotal colours}
Our procedure uses Carter's isophotal representation of the V frame. 
The successive contours at 0.1 mag intervals are fitted to each of the 
two frames to be compared, the mean surface brigtnesses calculated and the 
corresponding magnitudes and colours derived. A sliding mean smoothing is 
applied to the data for the outer contours and a graph of the colour 
against $\log r$ displayed. This might hopefully be linear, or nearly so, in 
the studied range. If it is not, {\em it is our practice to introduce 
corrections to the provisional values of the sky background for one or both 
of the frames,  
in order to get rid of the '' breaks" in the colour-radius relation 
typically associated with a poor choice of the background constant}. The reader 
is referred to the graphs published in Goudfrooij et al. (1994) for  
examples of such features. In Fig.1 we show a color profile with a rather 
important defect due to poor backgrounds, and its adopted correction.

The introduction of such ''aesthetic" corrections to the raw data might 
be criticized, since it assumes a regular behaviour of the colours at large 
$\log r$. This is however a reasonable hypothesis: the introduced 
corrections remain small, as shown by the statistics of Table 7. It should 
be noted that the {\em mean sky background values} derived for our large field 
frames are more precise than in previous works based on small field frames, 
where the sky was not reached at all. 
The problem lies in the presence of residual large-scale 
background fluctuations (see above): their effects are similar to those 
resulting from the poor evaluation of a constant background, and can be 
approximately corrected by the introduction of an ad hoc constant, or rather a 
set of constants, for the 5 frames in the colour set.  

The linear fit was finally performed on a range of $\log r$ selected so as to 
avoid the central regions affected by known dust patterns or possible residual 
seeing-induced errors, and the outermost regions visibly affected by deviations 
from the expected straight line.
 
\begin{table}
\caption[ ]{Corrections to provisional sky background values applied to cancel 
out ''breaks" in the run of colours aginst $\log r$. The table gives the mean 
absolute values of the applied corrections. 
Unit: \% of sky background.}
\begin{flushleft}
\begin{tabular}{llllll}
\hline      
								
Colour           &  U   &  B   &  V   &  R   &  I  \\	
\hline
 1st and 2nd run & 0.22 & 0.14 & 0.12 & 0.18 & 0.20 \\
         3rd run & 0.14 & 0.07 & 0.02 & 0.08 & 0.14 \\
\hline
\end{tabular}
\end{flushleft}
\end{table}

\begin{figure}
  \resizebox{\hsize}{!}{\includegraphics{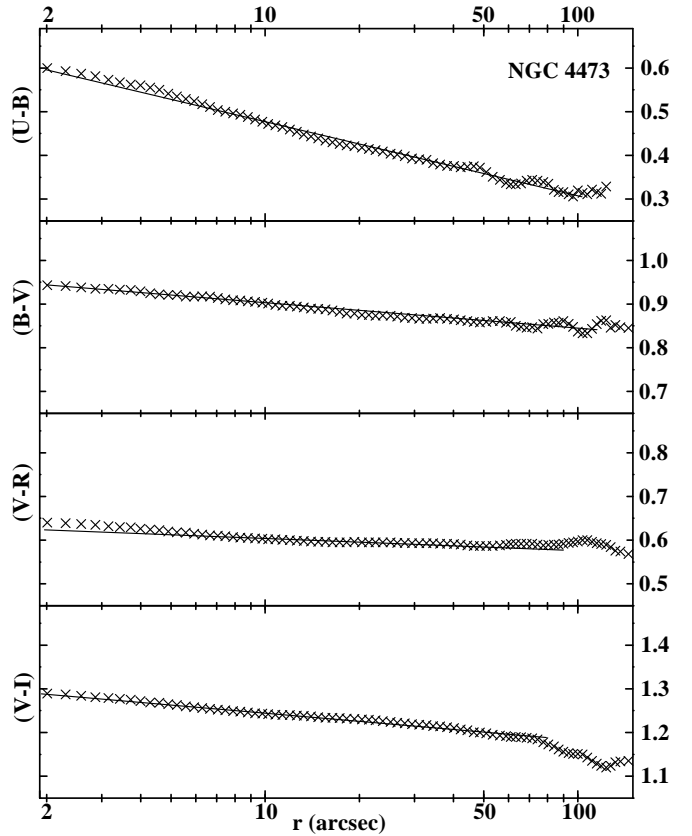}}
 \caption{Example of a set of ''regular" colour profiles for NGC4473. 
In this case, the colours are nearly linear in $\log r$ through the observed 
range. In this particular case a linear fit was made in the range 5--80 \arcsec 
to provide the tabulated data.}
 \label{ }
\end{figure}
\begin{figure}
  \resizebox{\hsize}{!}{\includegraphics{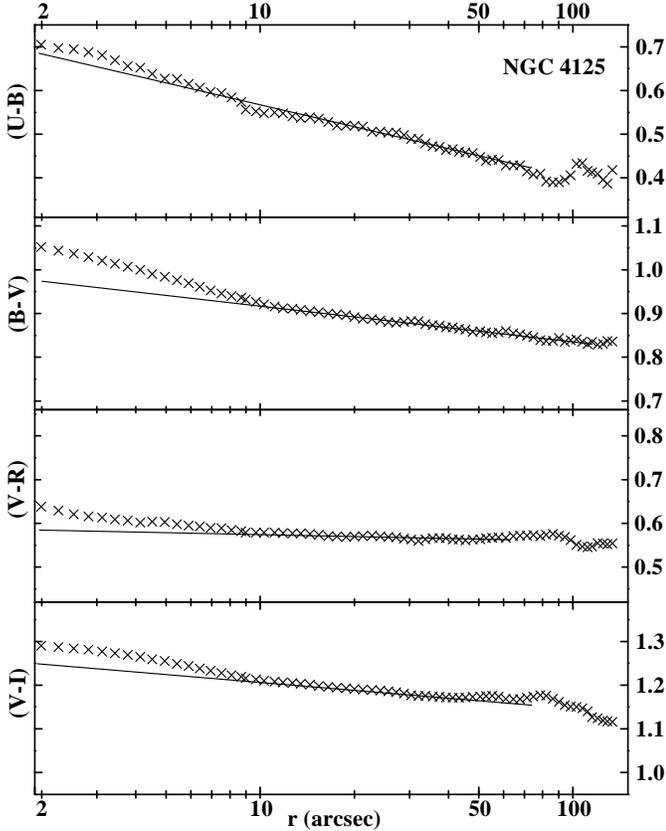}}
 \caption{Example of a set of colour profiles for NGC4125, a galaxy with 
a central dust pattern of importance index 3. 
In this case, the colours show a hump for $r < 10$ \arcsec, small in U$-$B but 
much larger in other colours. In this case, the linear fit was restricted to 
the range 10--80 \arcsec }
 \label{ }
\end{figure}
\begin{figure}
  \resizebox{\hsize}{!}{\includegraphics{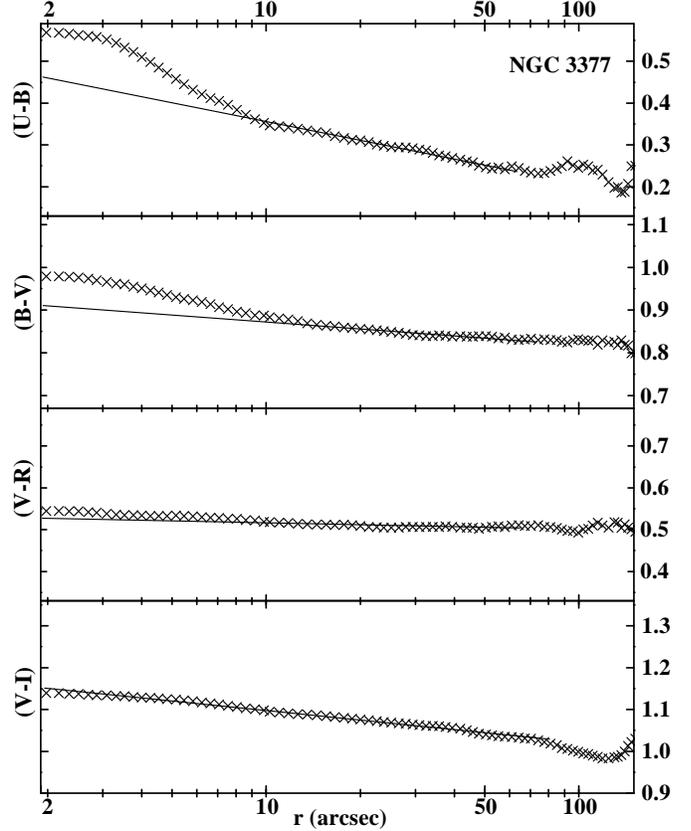}}
 \caption{Example of a set of colour profiles for NGC3377, a galaxy with a 
central red hump in U$-$B or B$-$V but not in V$-$I. This suggests a metallicity 
effect. The fit was obtained in the range 8--80 \arcsec.}
 \label{ }
\end{figure}

\subsubsection{Errors}

The noise is not a significant source of error in this type of work, because 
averages can be performed upon thousands of pixels in the galaxy regions of 
low S/N ratio. Residual noise effects at large $r$ can be easily recognized in 
sample plots of the data (see Fig. 1, 2 , 3 , 4).  
The main sources of errors lie:
\begin{itemize}
\item  in the calibration, giving a probable error $\sigma_C$, affecting equally 
all parts of a given object, but dependent upon the quality of the available 
aperture photometry in each colour. 

\item in the ''differential seeing", i.e. the small differences in PSF 
between the frames in a colour set, also after the adjustments described above. 
This error $\sigma_0$ occurs only near the center of the objects, say in a 
diameter of twice the seeing FWHM (it would be a much larger range without the 
performed adjustments).There is no obvious reason for it to be colour dependent.

\item in the sky background residual large-scale fluctuations, or equivalently, 
errors in the sky background estimates. This error $\sigma_S$ is strongly dependent
upon colour and the studied radius, or rather the corresponding surface brightness. 
\end{itemize}
These three components of the total error will now be considered, together with 
other relevant topics.

\begin{enumerate}
\item {\em Errors in colours from calibration inaccuracies.}

When a number of measurements are available for a given object, for instance 5 
apertures in PP88 and PN94, a probable error of the resulting 
calibration is readily derived from the dispersion of these measured values about 
the results of simulated aperture photometry, calculated from our data, i.e. 
V magnitudes, isophotal parameters and colours. For the preferred calibrations 
with Poulain's data, the computed error is often less than 0.01 in B$-$V or 
V$-$R but may rise to 0.02 in U$-$B or V$-$I. Still larger calibration errors,  
up to 0.04, have been estimated for objects with very scanty or uncertain
aperture photometry. These probable errors apply to the zero point of the colour 
regression as shown in Table 9, and to all colour data from the same object.    
  
\item {\em Residual errors from PSF equalization}.

The residual errors after this step in our data treatment can be objectively 
ascertained by comparing the ''central" B$-$R colours in our survey with 
the equivalent data from RM99, derived from high-resolution CFHT 
frames. For the present survey the ''central colours" are the integrated colours  
within a radius of 3 \arcsec. From RM99, Table 6, we find the colours at the 
isophote of 1.5 \arcsec, which are likely similar. The statistics of the 
difference 
B$-$R(new)--B$-$R(RM99) are for 31 objects in common: mean=0.01; $\sigma=0.038$. 
Assuming then that the errors are equal in the two surveys, the probable error 
associated with poor PSF adjustements is $\sigma_0=0.027$ in B$-$R. 
This source of error has no  
reason to vary significantly from one colour to another. 
It is independent of the error of calibration previously discussed. 

The comparison between the two surveys is made possible because the 
same set
of calibrations has been used, although the field of the CFHT frames was often
not sufficient to use all calibrations apertures. A minor part of the above
differences may come from this source. In RM99, the B$-$R of NGC2768 is quoted 
too red by 0.08 and that of NGC3610 too red by 0.10. 

\item {\em Errors from sky background inaccuracies}

Given $\epsilon_A$ and $\epsilon_C$, the relative errors in the sky background 
evaluation for frames A and C, A and C beeing a pair among UBVRI, the 
magnitude error in the colour A$-$C may be expressed as 
$\delta_{A-C}=-2.5 \log (1+\epsilon_A 10^{0.4\Delta \mu _A})+ 
2.5 \log (1+\epsilon_C 10^{0.4\Delta \mu _C})$.
Here $\Delta \mu _A$ is the magnitude contrast between the object and the sky 
in colour A. Using average values for the colours of E-galaxies and for the sky 
brightnesses, $\Delta \mu _A$ for any colour may be expressed in terms of 
$\Delta \mu _V$. Then, in the range of small $\delta_{A-C}$, the expression 
reduces to
$\delta_{A-C}= 1.0857(K_A \epsilon_A-K_C \epsilon_C) 10^{0.4\Delta \mu _V}$
where  $K_A=10^{0.4\Delta \mu _A}/10^{0.4\Delta \mu _V}$. 
  
$K_V=1$ by definition; we find $K_U=3.1$ and $K_I=1.55$, while $K_B$ and 
$K_R$ are slightly below 1.

The $\epsilon_A$,... are unknown, but it is feasible to get statistics of the 
linear combinations $K_A \epsilon_A-K_C \epsilon_C$. Indeed, as explained above, 
we have adopted ad hoc corrections to provisional sky background values, in order 
to regularise the colour-$\log r$ relations. It is reasonable to assume that the 
{\em errors left after these corrections are proportional to the adopted 
corrections themselves.} 
We take $\delta_{A-C}=\eta K_{AC}=\eta (K_A \epsilon_A-K_C \epsilon_C)$, 
where $\eta$ is a small constant and the $K_{AC}$ may be derived from the 
statistics of the adopted corrections given in Table 7. In practice somewhat 
different statistics have been calculated, to take into account the fact that 
our corrections for two colours are not necessarily uncorrelated. 
The constant $\eta$, different for our observing runs of 
2000 and 2001, was chosen so as to get a system of errors compatible 
with the appearance of the data and also with the errors found for the 
slopes of the colour-$\log r$ relation.    
The finally adopted errors $\sigma_S$ from sky background inaccuracies are 
given in Table 8. Note that this source of error is negligible for 
$\mu_V=22$ or smaller. Predicted errors are reduced for our {\em run 3} 
as compared to the two others.

\begin{table}
\caption[ ]{Probable errors $\sigma_S$ associated with sky background 
inaccuracies. 
Units: magnitude. Estimated errors are the same in V$-$R as in B$-$V.}
\begin{flushleft}
\begin{tabular}{llllll}
\hline      
								
Colour          & $\mu_V$ & U$-$B & U$-$V & B$-$V & V$-$I \\	
\hline
 1st and 2nd run & 23   & 0.023 & 0.026 & 0.008 & 0.012 \\
 id              & 24   & 0.058 & 0.064 & 0.020 & 0.030 \\
 id              & 24.5 & 0.093 & 0.102 & 0.032 & 0.048 \\
\hline
 3rd run & 23   & 0.017 & 0.019 & 0.006 & 0.010 \\
 id      & 24   & 0.042 & 0.048 & 0.015 & 0.026 \\
 id      & 24.5 & 0.067 & 0.077 & 0.024 & 0.041 \\
\hline
\end{tabular}
\end{flushleft}
\end{table}    
\begin{figure}
  \resizebox{\hsize}{!}{\includegraphics{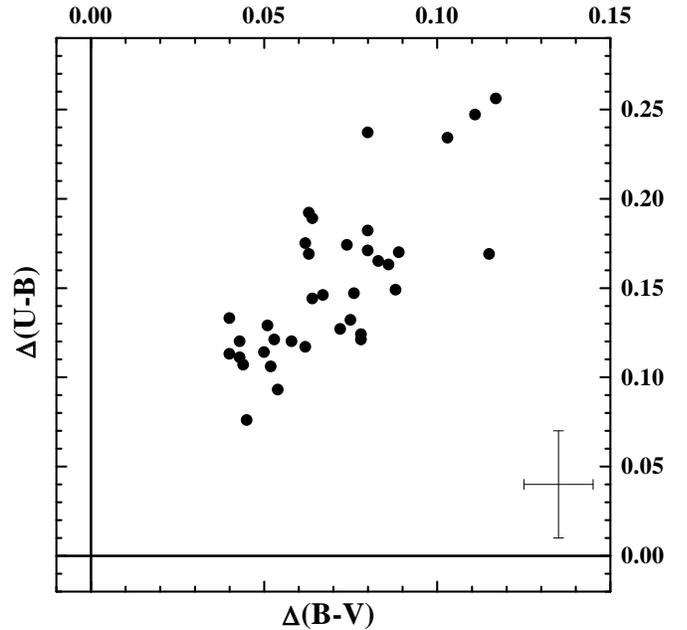}}
 \caption{Correlation between the colour gradients, {\em with the signs changed.} 
Abscissae: $\Delta_{BV}$. Ordinates: $\Delta_{UB}$. Compare with the 
similar diagram in RM00, or with the original 
correlation diagram between $\Delta_{BR}$ and $\Delta_{UR}$ in 
Peletier et al. (1990). The dispersion is clearly reduced here, which can only 
be attributed to an improved accuracy.}
 \label{ }
\end{figure}
\begin{figure}
  \resizebox{\hsize}{!}{\includegraphics{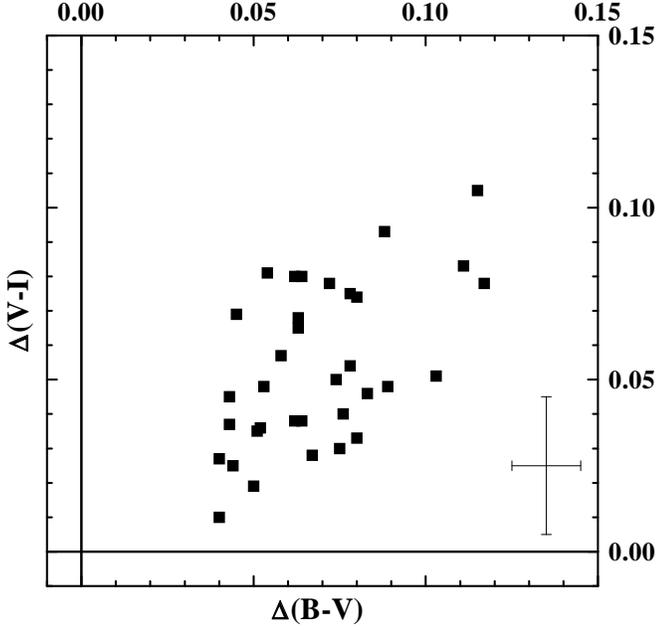}}
 \caption{Correlation between the colour gradients, {\em with the signs changed.}. 
Abscissae: $\Delta_{BV}$. Ordinates: $\Delta_{VI}$. Here the comparison with 
similar diagrams in RM00 does not show much increase in the accuracy of the new 
data. }
 \label{ }
\end{figure}
\item {\em Total errors}

The above estimated errors are independent and should be added quadratically. 
In the central region with $r<6$ \arcsec, the total error will be 
$\sigma_T = (\sigma_C^2+\sigma_0^2)^{1/2}$.
In the mean region with $18 < \mu_V < 22$ the total error equals the calibration 
error. Finally, in the outer range of $\mu_V > 23$ one can use 
$\sigma_T = (\sigma_C^2+\sigma_S^2)^{1/2}$.

\item {\em Errors from the corrections for red halo and other far wings effects} 

To our knowledge, the crossed convolutions used to correct for the red halo 
and similar effects do not give rise to random errors, but rather to  
systematic errors due to inaccuracies in the adopted PSFs. Such problems may be 
detected from the study of the distributions of the slopes of the 
colour-$\log r$ relations discussed below.  

\item {\em Errors in the slopes of the colour-$\log r$ relations}

These have been calculated by two complementary methods. On the one hand, 
we can look for the correlations between the slopes derived here and those from 
the literature, notably the data collected and discussed in RM00. 
Assuming then that the errors are the same in both sources, we get an estimate of 
our slope errors, hopefully an upper limit.
On the other hand we can consider the internal correlations between the slopes 
of the various colour-$\log r$ relations, specifically $\Delta_{UB}$ 
and others  
with $\Delta_{BV}$. Figure 5 and 6 shows the correlations of the 
U$-$B and V$-$I colour gradients with that in B$-$V. The coefficients of 
correlation are respectively 0.73 and 0.40. 
A weighted mean $Gm4$ of the slopes in 
the 4 colours has also been used as reference instead of $\Delta_{BV}$
with analogous results. 
From the dispersions of such correlations 
the slope errors can be estimated, if the error for the reference 
$\mathrm{d}{B-V}/d\log r$ or $Gm4$ is ''guessed". 
The two techniques give results in very good agreement, the internal correlations 
indicating somewhat smaller errors. 

{\em The probable errors of the slope estimates are then 0.03 in U$-$B or U$-$V, 
0.01 in B$-$V, 0.015 in V$-$R or B$-$R, 0.02 in V$-$I}.

\end{enumerate}

\begin{figure}
  \resizebox{\hsize}{!}{\includegraphics{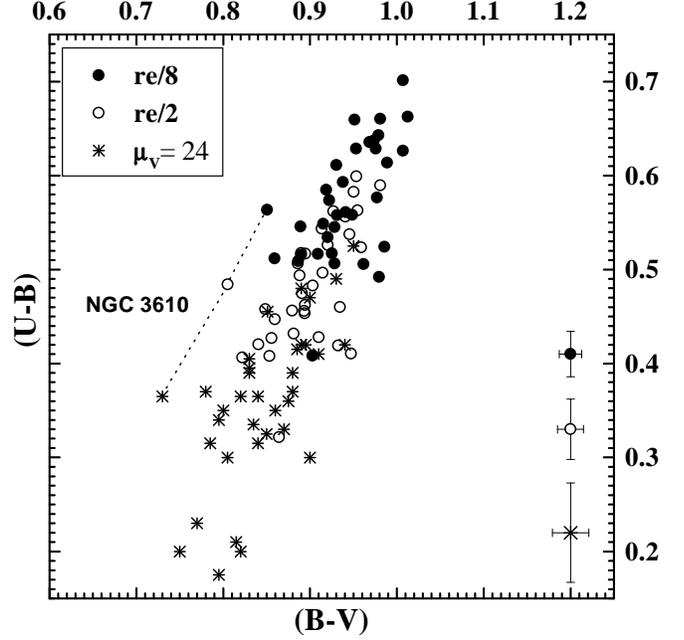}}
 \caption{Colour-colour diagram of U$-$B against B$-$V: the colours are 
calculated from the linear 
representations of Table 9 at the effective radius $r_e/8$ (dots), at 
$r_e/2$ (circles) and at the outermost range near $\mu_V=24$ (stars)
from Table 11. Although the various symbols refer to widely different 
regions of the galaxies they define a common relation.}
 \label{ }
\end{figure}
\begin{figure}
  \resizebox{\hsize}{!}{\includegraphics{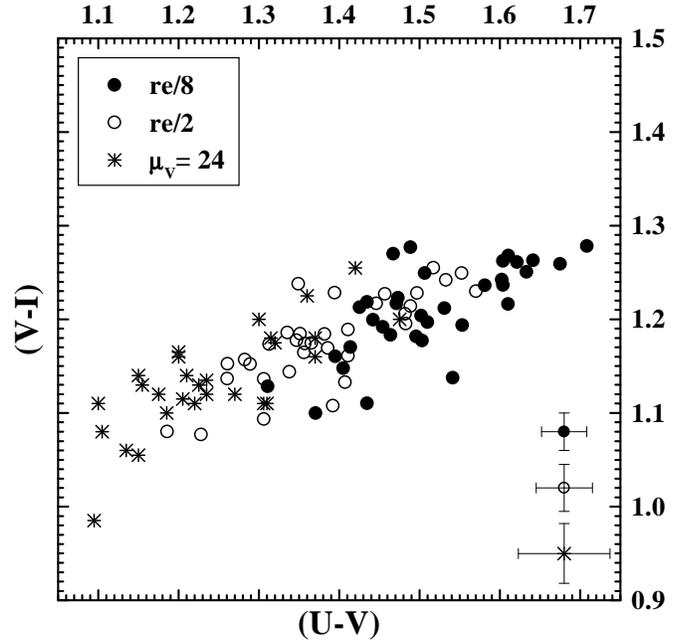}}
 \caption{Colour-colour diagram of V$-$I against U$-$V: the colours are 
calculated from the linear 
representations of Table 9 at the effective radius $r_e/8$ (dots), at 
$r_e/2$ (circles) and at the outermost range near $\mu_V=24$ (stars) from 
Table 11. }
 \label{ }
\end{figure}

\section{Observational results}
The available data from the present study are:
\begin{enumerate}
\item Table 9, giving for each object the linear representation of 
the colour-$\log r$ relations, i.e. the selected inner and outer radii 
of the fit, the zero point colour with its probable 
error $\sigma_C$ and the slope. The probable errors for the slopes are given 
above. This table allows easy calculation of the colours at any radius, 
notably the effective radius $r_e$ and others as considered below.

\item In electronic form only, the tables giving for each galaxy, as a function 
of the radius $r$, the V magnitude $\mu_V$ and the colours U$-$B, B$-$V, V$-$R, 
V$-$I. 
These tables are presently available with the U$-$B indices  
as observed, and consistent with Table 9. They will be later made available 
after correction for suspected effects of PSF far wings (see 
discussion in 3.2.5).
  
\item Series of graphs from the above tables, showing the colours as a function 
of $\log r$ or of $\mu_V$. Examples of these graphs are shown here to illustrate 
a number of properties of the radius-colour relations.

\item Table 10 gives the ''central'' colours according to several definitions:   
we have considered the colours integrated inside the area of radius $r=3$ \arcsec, 
and the colours calculated for $r=1.5$ \arcsec from the linear representations of 
Table 9. They should be nearly equal if these representations remain valid
at small $r$, which is not always the case: see 
below for a description of typical deviations.

\item Table 11 collects colours measured at the outermost range of the available 
data expressed in V magnitude. This gross limit varies between $\mu_V=23$ 
(for NGC4472) and 24.5. It is controlled by the size of the object and the ''cleanliness" 
of the nearby field. 
\end{enumerate}

{\em The SA0 galaxies NGC3115, 3607, 4550 and 5866 have been observed with 
the E-type sample. The corresponding results are given in the Tables, but they 
have been discarded from the discussion.}

\subsection{Statistical comparison with previous work}
\begin{itemize}
\item Assuming that colours result only from population variations, ''perfect" 
correlations between zero point colours in different pass-bands are expected.  
We have compared colour-colour correlations for previous surveys and the present 
one. For B$-$I against B$-$V we find from 
Goudfrooij et al. (1994) a coefficient of correlation $\rho=0.61$ (41 objects) 
compared to $\rho=0.94$ from Table 9 (37 objects). For U$-$R against B$-$R 
the data of Peletier et al. 1990 (38 objects) give $\rho=0.58$, while the 
present data leads to $\rho=0.94$. The improvement is less if we compare 
our results with  
the discussion in RM00 where the calibration of the available photometry 
was reconsidered. 
\item The correlation between colour gradients should also be as good as allowed 
by errors of measurements. Again we compare the 
correlation coefficients between gradients from the literature and our results 
for the same colours. From Goudfrooij et al. (1994), considering the gradient
$\Delta_{BI}$ against $\Delta_{BV}$, we find $\rho=0.78$ which is quite good.
The result derived from Table 9 is still better, i.e. $\rho=0.85$. The 
improvement is more pronounced when the $\Delta_{UR}$ against $\Delta_{BR}$ 
correlation is calculated from Peletier et al. (1990): one obtains 
$\rho=0.36$ only, instead of $\rho=0.88$ with our data. 
\footnote {The coefficients of correlation given here are of course different 
from the previously calculated ones in Sect. 3.2.7.6 , which refered to other 
colours with smaller gradients.}
\item Our colour measurements could be obtained at much lower surface brightness 
(or larger radii) than in previous work. The present data 
extend to $23.2 <\mu_V <25.2$, with a median value near 
$\mu_V=24.5$ in all colours. The tables by Goudfrooij et al. (1994), 
available from CDS Strasbourg, are mostly 
limited to $\mu_V=22.75$ in B$-$V and $\mu_V=22.25$ in V$-$I (median values), 
due to the small fields of the frames, notably in the I band 
(and probably an ad hoc cut off). 
Their published graphs generally extend to the same radius in both colours.  
The tables from Peletier et al. (1990), again at the CDS, extend to $\mu_V=23.2$ 
in B$-$R and $\mu_V=22.6$ in U$-$R (mean values) due to a systenatic cut-off at 
10\% of the sky. The printed tables may be still more severely truncated. 

{\em In summary, our colour data generally extends 1.5 to 2 magnitudes deeper than
in previous works}, so that ''external colours" refering to the level  
$\mu_V=24.5$ whenever possible, are presented in Table 11 with realistic error 
estimates.    
\end{itemize}
 
\subsection{Description of isophotal colour profiles.}
Most of the profiles relating colour to $\log r$, or equivalently to the 
surface brightness $\mu_V$, are {\em regular}, meaning that they deviate very 
little from a straight line in the range of abscissae relevant to the present data, 
roughly $r >2$ \arcsec and $\mu_V < 24.5$. In this case, the ''central colours" 
integrated within $r < 3$ \arcsec, and the colours calculated from the linear
representation at the average radius of $r=1.5$, differ very little. 
Central colour according to these two definitions are given in Table 10: 
compare columns (5) and (6) for U$-$V, and (3) with (7) for V$-$R.  
Figure 2 gives an example of a set of regular colour profiles for NGC4473. 
Of course the linear colour--$\log r$ relation is only approximate and breaks 
down at small $r$ for all galaxies observed at high resolution. Carollo et al. 
(1997) obtained V$-i$ maps of a number of E-galaxies from HST frames, 
disclosing minute colour structures in some cases. In RM99, 
the CFHT resolution proved sufficient for the detection and classification of 
B$-$R central '' red peaks". These are smoothed out, at least partly, with the OHP 
seeing. 

{\em Non regular} profiles have been observed in the following cases:
\begin{enumerate}
\item When an important dust pattern occurs near the center of an object, it 
produces a central red hump in the colour profile. This is the case for galaxies 
with the value of 3 for the {\em dust pattern importance index (DPII)} introduced 
in RM99, such as NGC2768, 4125 and 4374, but also for 5813 and 5831.  
The dust ring of NGC3607, type SA0, produces noticeable 
bumps in its colour profiles. Fig. 3 gives an example of the colour profiles 
of such a centrally dusty galaxy, i.e. NGC4125. In such cases, the integrated colours 
within $r=3$ \arcsec are redder than the extrapolated colours at $r=1.5$: this 
''extra reddening" is smaller in U$-$B than in other colours. The consideration of the
central reddening in V$-$R, a colour nearly insensitive to age-metallicity variations 
but sensitive to dust, gives a possibility to correct other colours for dust effects. 
This has been done, in some applications, for galaxies of {\em DPII} 3.   

\item A few objects show a central red hump in some of the colour profiles, 
specially U$-$B and also eventually B$-$V. This is the case of NGC3377, 3379, 
3610(?) and 4494. Fig.4 illustrates the case of NGC3377. For such objects the
''extra reddening" defined above is near zero in V$-$R and V$-$I. It is therefore
permitted to attribute it to a metallicity increase rather than to dust. 

\item On the contrary, the central colours U$-$B, B$-$V may be bluer than the 
extrapolation of the linear portion of the profile. This occurs for objects with 
larger than average colour gradients. The best such case is NGC4636; others are 
4283, 4478 and the SA0 4550. An ''extra blueing" in U$-$V is then apparent from the 
comparison of indices in columns (5)-(6) of Table 10. This is also marginally the 
case for such giant galaxies as NGC4406, 4472, 4649: according to RM99 the 
colour profile of such objects tend to flatten out near the center. 

\item NGC4486 is remarkable in showing a central ''blue deep", the U$-$B colour 
beeing 0.51 at $r=0$ as compared to 0.72 near $r= 6.5$ \arcsec. This central 
feature is probably somehow related to the famous non-thermal jet of this galaxy.  
The jet is of course very conspicuous in  
U$-$B , with a peak colour near -0.1. Needless to say, both the central 
''blue deep" 
and the jet are affected by seeing (and our attempts to correct its effects).
\end{enumerate}

\begin{figure}
  \resizebox{\hsize}{!}{\includegraphics{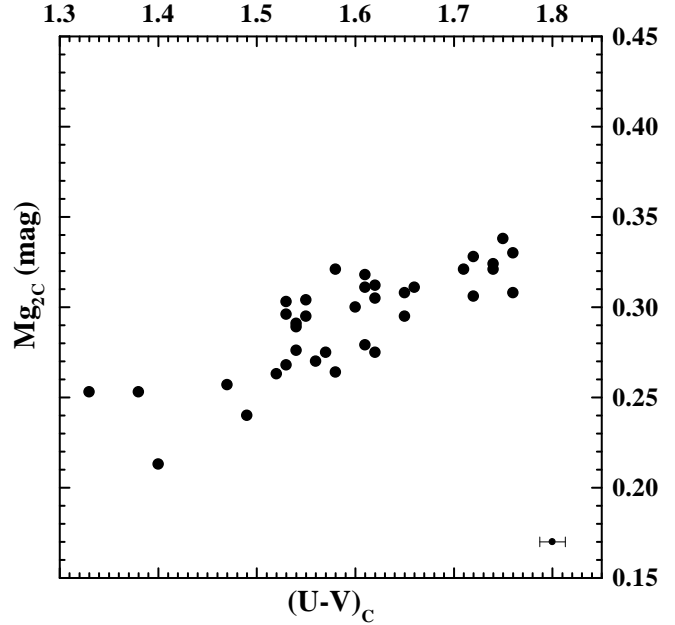}}
 \caption{Correlation between the near center U$-$V colour in abscissae, 
and the $Mg_2$ index from Faber et al. 1989, in ordinates}  
 \label{ }
\end{figure}
\begin{figure}
  \resizebox{\hsize}{!}{\includegraphics{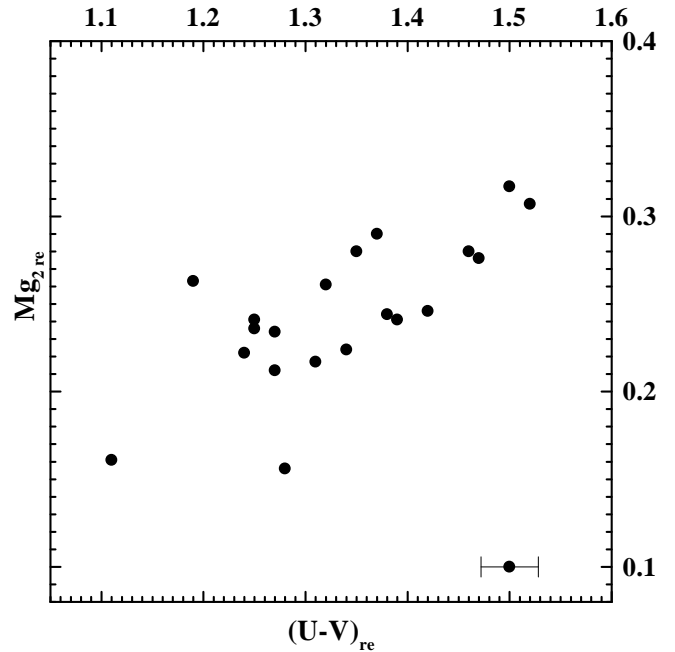}}
 \caption{Correlation between the U$-$V colour and the $Mg_2$ index at the 
effective radius. This index, and the $Mg_b$ one merged in the data, are 
taken from Kobayashi and Arimoto, 1999.}
 \label{ }
\end{figure}

\subsection{Correlations of interest.}
\subsubsection{Correlations between colour gradients.}
The correlations between the various colour gradients have already been 
noted as useful tools to evaluate the probable errors in gradients. 
These correlations are displayed in Fig. 5 and 6.

The coefficient of correlation between $\Delta_{UB}$ and $\Delta_{BV}$ 
is 0.75; that between $\Delta_{VI}$ and $\Delta_{BV}$ is only 0.40. 
For $\Delta_{VR}$ it falls to 0.20, because the errors are of the same order 
of magnitude as the V$-$R gradients.    
Imposing regression lines {\em running through the origin}, the relative slopes 
are $<\Delta_{UB}/\Delta_{BV}=2.2>$, $<\Delta_{VI}/\Delta_{BV}=0.8>$ and 
$<\Delta_{VR}/\Delta_{BV}=0.25>$. These relative slopes are in good agreement 
with the results in RM00, and its conclusion, i.e. the negligible influence of 
dust upon colour gradient, is confirmed. 

The distribution of colour gradients for E-galaxies may be of interest. The 
following parameters are found: $<\Delta_{UB}>=-0.152$ with $\sigma=0.044$; 
$<\Delta_{BV}>=-0.070$ with $\sigma=0.021$; $<\Delta_{VR}>=-0.018$ with 
$\sigma=0.012$; $<\Delta_{VI}>=-0.054$ with $\sigma=0.023$. The dispersions are 
not much larger than the errors estimated above. The distributions are 
asymmetric: there are 4 objects with $\Delta_{BV}$ larger than $+1.8\sigma$ 
above the mean, but none at less than the same deviation. These       
galaxies, with $\Delta_{BV}$ clearly steeper than average, by about twice the 
estimated probable error of measurement, 
are NGC4283, 4478, 4564 and 4636, seemingly a random collection. 

Remark: An attempt to sort the E-galaxies by flattening as measured in  
MM94, and to look for some relation to the gradients, 
lead to negative results. Similarly no significant difference was found 
between diE and other galaxies. 

The SA0 NGC4550 has quite exceptional gradients in all colours, and  
an admixture of dust and relatively young stars could be invoked to explain its
properties. This object also has very remarkable kinematics, as first described 
by Rubin et al. (1992); a model has been proposed by Rix et al. (1992).  
The few other S0s in the present sample are similar to Es with 
regard to their colour gradients.

\subsubsection{Colour-colour diagrams}

Many colour-colour diagrams can be built from the present data. 
The indices used  
may be calculated from the linear representations in Table 9 at the effective radius 
$r_e$, at a near center position $r_e/8$, and at an intermediate position $r_e/2$. 
The system of $r_e$ here used is an average of estimates in MM94, Prugniel 
and H\'eraudeau, 1998, and the RC3. 
It is satisfactory that the various indices $\mathrm{(B-V)}_{r_e}$, 
$\mathrm{(B-V)}_{r_e/2}$, $\mathrm{(B-V)}_{r_e/8}$, define a common diagram 
with the corresponding U$-$B. This may suggest that a common physical variable 
controls the variations inside an object, and the object to object changes, 
of the two colours. Such graphs readily show larger than average 
calibrations errors:  
for instance, the U$-$B colour of NGC3610 is clearly too red for its B$-$V. 

We have also traced colour-colour diagrams for the ''central" colours, 
(Table 10), i.e.  
integrated in the radius $r < 3$ \arcsec. They are similar to those traced with 
the interpolated colours, but with larger dispersions: this is not surprising 
since the central colours suffer from larger errors (see 3.2.7).  

Finally, one can trace colour-colour diagrams with the ''outermost colours" 
collected in Table 11. They are in fair agreement with the diagrams derived from 
Table 9, and extend these towards the blue. We show in Fig. 7 a composite 
colour-colour diagram U$-$B against B$-$V, using the colours ar $r_e/8$, $r_e/2$ 
and the outermost range. Similarly Fig. 8 displays the diagram of V$-$I 
against U$-$V.

\begin{table*}
\caption[ ]{Linear representation of colour against $\log r$. Successive 
columns: NGC No.; Type; $r_i$ inner radius of calculation; $r_o$ outer radius 
of calculation; $r_0$ radius of colour evaluation (in log and "); 
U$-$B at $r_0$ and estimated standard error; $\Delta_{UB}$ radial gradient; 
B$-$V at $r_0$ and estimated standard error; $\Delta_{BV}$ radial gradient; 
V$-$R at $r_0$ and estimated standard error; $\Delta_{VR}$ radial gradient; 
V$-$I at $r_0$ and estimated standard error; $\Delta_{VI}$ radial gradient; 
$du$ Dust visibility index (Michard 1999)
Notes: 2974: V$-$R and V$-$I not measurable; 3193: V$-$I not measurable.  }
\begin{flushleft}
\begin{tabular}{llllllllllllll}
\hline

NGC  & Type & $r_i$ & $r_o$ & $r_0$ & U$-$B & $\Delta_{UB}$ & B$-$V  & $\Delta_{BV}$ & V$-$R & $\Delta_{VR}$ & V$-$I  & $\Delta_{VI}$ & 
$du$ \\
\hline
2768 & diE & 10 & 80 & 1.467 & 0.521$\pm.03$ & -0.093 & 0.892$\pm.02$ & -0.054 & 0.536$\pm.02$ & -0.012 & 1.136$\pm.02$ & -0.081 & 3 \\
2974 & diE & 8 & 80 & 1.404 & 0.429$\pm.01$ & -0.237 & 0.905$\pm.01$ & -0.080 & - & - & - & - & 3 \\
3115 & SA0 & 8 & 100 & 1.464 & 0.501$\pm.01$ & -0.169 & 0.912$\pm.01$ & -0.069 & 0.585$\pm.01$ & -0.010 & 1.222$\pm.01$ & -0.047 & 0 \\
3193 & unE & 6 & 40 & 1.169 & 0.461$\pm.02$ & -0.163 & 0. 924$\pm.01$ & -0.086 & 0.567$\pm.02$ & 0.009 & - & - &  0 \\
3377 & diE & 8 & 80 & 1.399 & 0.298$\pm.02$ & -0.144 & 0.854$\pm.01$ & -0.064 & 0.510$\pm.01$ & -0.014 & 1.067$\pm.01$ & -0.080 & 1- \\
3377 & diE & 8 & 80 & 1.403 & 0.289$\pm.02$ & -0.170 & 0.852$\pm.01$ & -0.075 & 0.505$\pm.01$ & -0.028 & 1.074$\pm.01$ & -0.105 & 1- \\
3379 & unE & 8 & 100 & 1.460 & 0.538$\pm.02$ & -0.111 & 0.918$\pm.01$ & -0.043 & 0.585$\pm.01$ & -0.013 & 1.206$\pm.01$ & -0.037 & 1- \\
3605 & boE & 4 & 30 & 1.043 & 0.403$\pm.03$ & -0.175 & 0.821$\pm.02$ & -0.062 & 0.517$\pm.02$ & -0.014 & 1.076$\pm.02$ & -0.038 &  0 \\
3607 & SA0 & 15 & 100 & 1.569 & 0.451$\pm.02$ & -0.111 & 0.882$\pm.01$ & -0.067 & 0.537$\pm.02$ & -0.017 & 1.149$\pm.02$ & -0.046 & 3 \\
3608 & boE & 5 & 50 & 1.201 & 0.415$\pm.03$ & -0.189 & 0.949$\pm.01$ & -0.064 & 0.542$\pm.01$ & -0.020 & 1.175$\pm.01$ & -0.038 & 1- \\
3610 & diE & 4 & 50 & 1.159 & 0.448$\pm.04$ & -0.132 & 0.785$\pm.01$ & -0.075 & 0.503$\pm.01$ & 0.003 & 1.144$\pm.02$ & -0.030 & 0 \\
3613 & diE & 6 & 60 & 1.287 & 0.470$\pm.03$ & -0.174 & 0.875$\pm.02$ & -0.074 &0.533$\pm.04$ & -0.018 & 1.100$\pm.04$ & -0.050 & 1- \\
3640 & boE & 5 & 60 & 1.266 & 0.455$\pm.03$ & -0.120 & 0.892$\pm.02$ & -0.043 & 0.544$\pm.02$ & -0.012 & 1.162$\pm.03$ & -0.045 & 0 \\
3872 & diEp & 5 & 60 & 1.202 & 0.506$\pm.03$ & -0.127 & 0.928$\pm.02$ & -0.072 & 0.571$\pm.01$ & -0.005 & 1.176$\pm.02$ & -0.078 & 0 \\
4125 & diE & 10 & 80 & 1.445 & 0.497$\pm.03$ & -0.165 & 0.890$\pm.02$ & -0.083 & 0.568$\pm.03$ & -0.011 & 1.183$\pm.03$ & -0.046 & 3 \\ 
4261 & boE & 5 & 80 & 1.320 & 0.588$\pm.03$ & -0.170 & 0.948$\pm.01$ & -0.089 & 0.585$\pm.01$ & -0.014 & 1.246$\pm.02$ & -0.048 & 1- \\
4278 & diE & 10 & 60 & 1.361 & 0.408$\pm.02$ & -0.149 & 0.845$\pm.01$ & -0.088 & 0.554$\pm.03$ & -0.031 & 1.145$\pm.03$ & -0.093 &  ? \\
4283 & unE & 5 & 30 & 1.080 & 0.410$\pm.02$ & -0.256 & 0.870$\pm.01$ & -0.117 & 0.533$\pm.02$ & -0.040 & 1.147$\pm.02$ & -0.078 & 0 \\
4365 & boE & 5 & 120 & 1.430 & 0.552$\pm.02$ & -0.120 & 0.939$\pm.01$ & -0.058 & 0.588$\pm.01$ & -0.011 & 1.226$\pm.02$ & -0.057 & 0 \\
4374 & unE & 8 & 80 & 1.394 & 0.497$\pm.03$ & -0.107 & 0.915$\pm.01$ & -0.044 & 0.565$\pm.02$ & -0.020 & 1.189$\pm.02$ & -0.025 & 3 \\
4387 & boE & 5 & 30 & 1.102 & 0.431$\pm.02$ & -0.106 & 0.883$\pm.01$ & -0.052 & 0.565$\pm.01$ & -0.028 & 1.170$\pm.02$ & -0.036 & 0 \\
4406 & boE & 5 & 80 & 1.312 & 0.463$\pm.02$ & -0.121 & 0.961$\pm.01$ & -0.078 & 0.561$\pm.01$ & -0.018 & 1.204$\pm.01$ & -0.054 & 0 \\
4472 & unE & 5 & 150 & 1.510 & 0.590$\pm.02$ & -0.133 & 0.963$\pm.01$ & -0.040 & 0.592$\pm.01$ & -0.005 & 1.257$\pm.01$ & -0.010 & 0 \\
4473 & diE & 5 & 80 & 1.310 & 0.423$\pm.03$ & -0.169 & 0.881$\pm.01$ & -0.063 & 0.598$\pm.01$ & -0.021 & 1.225$\pm.01$ & -0.065 & 0 \\
4478 & boE & 6 & 50 & 1.286 & 0.300$\pm.02$ & -0.234 & 0.806$\pm.01$ & -0.103 & 0.525$\pm.01$ & -0.023 & 1.129$\pm.01$ & -0.051 & 0 \\
4486 & unE & 10 & 120 & 1.559 & 0.566$\pm.02$ & -0.192 & 0.921$\pm.01$ & -0.063 & 0.601$\pm.01$ & -0.019 & 1.235$\pm.02$ & -0.068 & 0 \\
4494 & unE & 5 & 80 & 1.340 & 0.452$\pm.03$ & -0.114 & 0.862$\pm.01$ & -0.050 & 0.518$\pm.02$ & -0.016 & 1.137$\pm.02$ & -0.019 & ? \\
4550 & SA0 & 5 & 40 & 1.199 & 0.260$\pm.02$ & -0.252 & 0.826$\pm.01$ & -0.196 & 0.511$\pm.01$ & -0.108 & 1.119$\pm.02$ & -0.185 & 0 \\
4551 & boE & 5 & 40 & 1.184 & 0.475$\pm.02$ & -0.113 & 0.880$\pm.01$ & -0.040 & 0.525$\pm.02$ & -0.014 & 1.151$\pm.02$ & -0.027 & 0 \\
4552 & unE & 5 & 100 & 1.366 & 0.489$\pm.01$ & -0.171 & 0.943$\pm.01$ & -0.080 & 0.546$\pm.01$ & -0.035 & 1.191$\pm.02$ & -0.074 & 0 \\
4564 & diE & 5 & 50 & 1.233 & 0.369$\pm.02$ & -0.247 & 0.884$\pm.01$ & -0.111 & 0.545$\pm.01$ & -0.024 & 1.124$\pm.02$ & -0.083 & 1- \\
4621 & diE & 5 & 80 & 1.327 & 0.522$\pm.02$ & -0.182 & 0.919$\pm.01$ & -0.080 & 0.584$\pm.01$ & -0.004 & 1.216$\pm.02$ & -0.033 & 0 \\
4636 & unE & 7 & 100 & 1.416 & 0.495$\pm.02$ & -0.169 & 0.906$\pm.02$ & -0.115 & 0.562$\pm.02$ & -0.034 & 1.185$\pm.03$ & -0.105 & ? \\ 
4649 & unE & 5 & 150 & 1.515 & 0.592$\pm.01$ & -0.121 & 0.982$\pm.01$ & -0.053 & 0.571$\pm.01$ & -0.036 & 1.231$\pm.02$ & -0.048 & 0 \\
5322 & boE & 5 & 80 & 1.313 & 0.445$\pm.03$ & -0.146 & 0.843$\pm.02$ & -0.067 & 0.530$\pm.02$ & -0.001 & 1.091$\pm.03$ & -0.028 & 0 \\
5576 & boEp & 5 & 60 & 1.216 & 0.382$\pm.02$ & -0.147 & 0.821$\pm.01$ & -0.076 & 0.521$\pm.01$ & -0.003 & 1.126$\pm.01$ & -0.040 & 0 \\
5813 & unE & 5 & 60 & 1.229 & 0.477$\pm.01$ & -0.076 & 0.945$\pm.01$ & -0.045 & 0.589$\pm.01$ & -0.022 & 1.244$\pm.02$ & -0.069 & 1 \\
5831 & diE & 4 & 50 & 1.175 & 0.423$\pm.03$ & -0.124 & 0.876$\pm.02$ & -0.078 & 0.515$\pm.02$ & -0.040 & 1.168$\pm.02$ & -0.075 & ? \\
5846 & unE & 5 & 100 & 1.407 & 0.594$\pm.02$ & -0.129 & 0.955$\pm.02$ & -0.051 & 0.586$\pm.02$ & -0.016 & 1.245$\pm.02$ & -0.035 & 0 \\
5866 & SA0 & 15 & 110 & 1.612 & 0.367$\pm.02$ & -0.190 & 0.819$\pm.01$ & -0.100 & 0.537$\pm.01$ & -0.048 & 1.095$\pm.02$ & -0.042 & 3+ \\
5982 & boE & 8 & 100 & 1.447 & 0.431$\pm.02$ & -0.117 & 0.868$\pm.02$ & -0.062 & 0.522$\pm.02$ & -0.043 & 1.145$\pm.02$ & -0.080 & ? \\
\hline
\end{tabular}
\end{flushleft}
\end{table*}
 
\subsubsection{The U$-$V colour as a metallicity index for ellipticals ?}

Burstein et al. (1988) showed a correlation between the central $Mg_2$ index 
and a global B$-$V colour measured in a large aperture (see also Bender et al. ,
1993). This type of correlation is reconsidered here using the U$-$V colour 
which is much more sensitive to metallicity than B$-$V, and taking advantage of 
recent estimates of the $Mg_2$ index far from center.  

Two correlations between the $Mg_2$ index and U$-$V are 
considered in Fig. 9 and 10. The first shows the relation between the two 
quantities near the galaxy center: the $Mg_{2C}$ index is taken from the tabulation
by Faber et al. 1989. The $\mathrm{(U-V)}_C$ index is the integrated colour in a 
circle of radius $r=3$ \arcsec. The value for three galaxies with important 
central dust patterns have been corrected by reference to the central 
bump in (V$-$R), a colour sensitive to dust but less so to metallicity changes. 
The coefficient of correlation reaches 0.825. Taking $\mathrm{(U-V)}_C$ as $x$ and 
$Mg_{2C}$ as $y$ we find the regression $y=0.221\pm.027x-0.061\pm.003$.  

The second, i.e. Fig. 10, displays the correlation between the U$-$V colour and 
the $Mg_2$ index 
at the effective radius $r_e$. $Mg_{2 re}$ has been taken from Kobayashi and 
Arimoto (1999) (KA99). To increase the number of data points the $Mg_b$ gradients were  
introduced, using the linear relation $Mg_b\,=15\,Mg_2$ derived from the 
distribution of the values of both indices in the tables of KA99. 
The coefficient of correlation still reaches 0.72. Again, with the colour in $x$ 
and the $Mg_{2 re}$ in $y$ we find $y=0.283\pm.069x-0.133\pm.007$. The difference 
to the above correlation for central indices is barely significant.   
 
The quality of these correlations proves that both indices are essentially controlled 
by the same physical variables, and leaves little room for the effects of 
diffuse dust upon the colours of E-galaxies.

\begin{table}
\caption[ ]{Columns (1) to (5): integrated colours within a radius $r=3$ \arcsec, 
i.e. $UB_3$, $BV_3$ , $VR_3$ , $VI_3$, $UV_3$; (6): $UV_{1.5}$
U$-$V colour at $r=1.5\arcsec$  calculated from Table 9 . (7) $VR_{1.5}$
V$-$R colour at $r=1.5\arcsec$ }
\begin{flushleft}
\begin{tabular}{llllllllll}
\hline 
NGC  & $UB_3$ & $BV_3$ & $VR_3$ & $VI_3$ & $UV_3$ & 
$UV_{1.5}$ & $VR_{1.5}$  \\
\hline 
2768 & 0.72 & 1.04 & 0.61 & 1.28 & 1.76 & 1.60 & 0.55 \\  
2974 & 0.58 & 1.03 &  -   &  -   & 1.60 & 1.72 & 0.00 \\
3115 & 0.68 & 1.03 & 0.60 & 1.25 & 1.72 & 1.72 & 0.60 \\ 
3193 & 0.60 & 0.96 & 0.65 & 1.32 & 1.55 & 1.63 & 0.58 \\
3377 & 0.56 & 0.98 & 0.54 & 1.14 & 1.54 & 1.41 & 0.53 \\
3377 & 0.60 & 0.98 & 0.55 & 1.14 & 1.57 & 1.44 & 0.54 \\
3379 & 0.72 & 1.03 & 0.60 & 1.23 & 1.76 & 1.65 & 0.60 \\ 
3605 & 0.50 & 0.90 & 0.57 & 1.13 & 1.40 & 1.43 & 0.53 \\
3607 & 0.63 & 1.04 & 0.64 & 1.31 & 1.67 & 1.69 & 0.56 \\
3608 & 0.61 & 1.01 & 0.60 & 1.24 & 1.62 & 1.62 & 0.56 \\ 
3610 & 0.60 & 0.88 & 0.55 & 1.07 & 1.47 & 1.44 & 0.50 \\
3613 & 0.63 & 0.98 & 0.57 & 1.21 & 1.61 & 1.63 & 0.53 \\
3640 & 0.58 & 0.94 & 0.59 & 1.21 & 1.52 & 1.52 & 0.56 \\
3872 & 0.63 & 0.98 & 0.62 & 1.22 & 1.61 & 1.64 & 0.58 \\
4125 & 0.73 & 1.06 & 0.65 & 1.29 & 1.79 & 1.64 & 0.63 \\
4261 & 0.73 & 1.03 & 0.61 & 1.29 & 1.76 & 1.83 & 0.60 \\
4278 & 0.55 & 0.99 & 0.60 & 1.21 & 1.54 & 1.53 & 0.64 \\
4283 & 0.56 & 0.96 & 0.60 & 1.18 & 1.53 & 1.62 & 0.57 \\
4365 & 0.71 & 1.04 & 0.61 & 1.28 & 1.74 & 1.71 & 0.64 \\
4374 & 0.63 & 1.03 & 0.64 & 1.30 & 1.66 & 1.60 & 0.59 \\
4387 & 0.51 & 0.98 & 0.57 & 1.16 & 1.49 & 1.46 & 0.59 \\
4406 & 0.59 & 1.02 & 0.57 & 1.26 & 1.61 & 1.65 & 0.58 \\
4472 & 0.70 & 1.02 & 0.61 & 1.29 & 1.72 & 1.78 & 0.60 \\
4473 & 0.60 & 0.95 & 0.64 & 1.29 & 1.55 & 1.57 & 0.62 \\
4478 & 0.45 & 0.89 & 0.55 & 1.17 & 1.33 & 1.48 & 0.55 \\
4486 & 0.55 & 0.98 & 0.63 & 1.26 & 1.53 & 1.84 & 0.63 \\
4494 & 0.65 & 0.93 & 0.57 & 1.21 & 1.57 & 1.50 & 0.54 \\
4550 & 0.39 & 0.95 & 0.58 & 1.22 & 1.34 & 1.54 & 0.62 \\
4551 & 0.59 & 0.99 & 0.58 & 1.23 & 1.58 & 1.51 & 0.54 \\
4552 & 0.67 & 1.07 & 0.61 & 1.29 & 1.74 & 1.73 & 0.59 \\
4564 & 0.63 & 0.95 & 0.61 & 1.19 & 1.58 & 1.63 & 0.57 \\
4621 & 0.72 & 1.00 & 0.63 & 1.27 & 1.72 & 1.74 & 0.59 \\
4636 & 0.67 & 1.00 & 0.60 & 1.29 & 1.66 & 1.79 & 0.61 \\
4649 & 0.71 & 1.04 & 0.62 & 1.29 & 1.75 & 1.81 & 0.62 \\
5322 & 0.62 & 0.91 & 0.62 & 1.21 & 1.54 & 1.53 & 0.53 \\
5576 & 0.53 & 0.85 & 0.58 & 1.17 & 1.38 & 1.43 & 0.53 \\
5813 & 0.64 & 1.01 & 0.65 & 1.33 & 1.65 & 1.55 & 0.61 \\
5831 & 0.59 & 0.95 & 0.57 & 1.25 & 1.54 & 1.50 & 0.55 \\
5846 & 0.69 & 1.01 & 0.62 & 1.28 & 1.71 & 1.77 & 0.61 \\
5866 & -    &  -   &  -   &  -   &  -   & 1.60 & 0.61 \\
5982 & 0.59 & 0.94 & 0.54 & 1.18 & 1.53 & 1.53 & 0.58 \\
\hline      
								
\hline
\end{tabular}
\end{flushleft}
\end{table}
     
\begin{table}
\caption[ ]{''External colours" measured in the outermost range of the 
data. Successive columns: $\mu_V$ V surface brightness of measurement, a range of 0.5 
magnitude centered at the quoted $\mu_V$ beeing used. Columns (2), (3), 
(4) and (5): U$-$B, B$-$V, V$-$R, V$-$I respectively with estimated errors.
These are the same in V$-$R as in B$-$V. }

\begin{flushleft}
\begin{tabular}{llllll}
\hline 
NGC  & $\mu_V$ & U$-$B & B$-$V & V$-$R & V$-$I \\
\hline       
2768 & 24.25 & 0.455$\pm.06$ & 0.851$\pm.02$ & 0.530 & 1.110$\pm.04$ \\ 
2974 & id    & 0.335$\pm.06$ & 0.835$\pm.03$ & 0.000 & 0.000$\pm.00$ \\ 
3115 & id    & 0.365$\pm.05$ & 0.855$\pm.02$ & 0.580 & 1.180$\pm.03$ \\ 
3193 & 23.5  & 0.370$\pm.04$ & 0.880$\pm.02$ & 0.000 & 0.000$\pm.00$ \\ 
3377 & 24.5  & 0.210$\pm.06$ & 0.815$\pm.02$ & 0.505 & 1.000$\pm.03$ \\ 
3377 & 24.25 & 0.175$\pm.05$ & 0.795$\pm.02$ & 0.480 & 0.950$\pm.03$ \\ 
3379 & 24.0  & 0.480$\pm.04$ & 0.890$\pm.02$ & 0.575 & 1.160$\pm.03$ \\ 
3605 & 23.5  & 0.340$\pm.03$ & 0.795$\pm.02$ & 0.510 & 1.060$\pm.02$ \\ 
3607 & 24.25 & 0.370$\pm.07$ & 0.845$\pm.03$ & 0.530 & 1.130$\pm.04$ \\ 
3608 & 23.75 & 0.300$\pm.05$ & 0.900$\pm.02$ & 0.525 & 1.165$\pm.03$ \\ 
3610 & 24.25 & 0.365$\pm.05$ & 0.730$\pm.02$ & 0.510 & 0.985$\pm.04$ \\ 
3613 & 24.5  & 0.395$\pm.07$ & 0.830$\pm.02$ & 0.500 & 1.130$\pm.04$ \\ 
3640 & 23.5  & 0.360$\pm.04$ & 0.875$\pm.02$ & 0.540 & 1.135$\pm.03$ \\ 
3872 & 24.5  & 0.420$\pm.07$ & 0.890$\pm.03$ & 0.555 & 1.110$\pm.05$ \\ 
4125 & 23.5  & 0.405$\pm.03$ & 0.830$\pm.01$ & 0.550 & 1.120$\pm.02$ \\ 
4278 & 23.5  & 0.350$\pm.04$ & 0.800$\pm.02$ & 0.530 & 1.055$\pm.03$ \\ 
4283 & 23.75 & 0.300$\pm.05$ & 0.805$\pm.02$ & 0.510 & 1.080$\pm.03$ \\ 
4365 & 24.   & 0.470$\pm.04$ & 0.900$\pm.02$ & 0.585 & 1.180$\pm.03$ \\ 
4374 & 23.25 & 0.420$\pm.03$ & 0.895$\pm.01$ & 0.545 & 1.180$\pm.03$ \\ 
4387 & 24.5  & 0.350$\pm.08$ & 0.860$\pm.03$ & 0.550 & 1.140$\pm.05$ \\ 
4406 & 23.25 & 0.330$\pm.04$ & 0.870$\pm.02$ & 0.565 & 1.160$\pm.03$ \\ 
4472 & 23.   & 0.490$\pm.03$ & 0.930$\pm.01$ & 0.590 & 1.255$\pm.02$ \\ 
4473 & 24.5  & 0.315$\pm.08$ & 0.840$\pm.03$ & 0.575 & 1.130$\pm.05$ \\ 
4478 & 24.   & 0.200$\pm.04$ & 0.750$\pm.02$ & 0.490 & 1.090$\pm.03$ \\ 
4486 & 23.5  & 0.415$\pm.03$ & 0.885$\pm.02$ & 0.590 & 1.200$\pm.03$ \\ 
4494 & 24.5  & 0.390$\pm.08$ & 0.830$\pm.03$ & 0.500 & 1.110$\pm.05$ \\ 
4550 & 24.0  & 0.120$\pm.06$ & 0.700$\pm.02$ & 0.470 & 1.010$\pm.03$ \\ 
4551 & 24.5  & 0.325$\pm.07$ & 0.850$\pm.03$ & 0.520 & 1.120$\pm.04$ \\ 
4552 & 24.   & 0.390$\pm.06$ & 0.880$\pm.02$ & 0.515 & 1.120$\pm.03$ \\ 
4564 & 24.5  & 0.200$\pm.07$ & 0.820$\pm.02$ & 0.530 & 1.020$\pm.03$ \\ 
4621 & 24.   & 0.390$\pm.06$ & 0.370$\pm.02$ & 0.565 & 1.175$\pm.03$ \\ 
4636 & 23.5  & 0.370$\pm.03$ & 0.780$\pm.02$ & 0.510 & 1.140$\pm.03$ \\ 
4649 & 23.5  & 0.525$\pm.04$ & 0.950$\pm.02$ & 0.540 & 1.200$\pm.03$ \\ 
5322 & 24.25 & 0.315$\pm.07$ & 0.785$\pm.03$ & 0.520 & 1.110$\pm.04$ \\ 
5576 & 23.75 & 0.230$\pm.05$ & 0.770$\pm.02$ & 0.525 & 1.085$\pm.03$ \\ 
5813 & 24.   & 0.410$\pm.06$ & 0.910$\pm.02$ & 0.575 & 1.175$\pm.03$ \\ 
5831 & 23.   & 0.365$\pm.06$ & 0.840$\pm.02$ & 0.495 & 1.115$\pm.04$ \\ 
5846 & 24.   & 0.420$\pm.06$ & 0.940$\pm.02$ & 0.580 & 1.225$\pm.03$ \\ 
5866 & 24.5  & 0.270$\pm.08$ & 0.775$\pm.03$ & 0.505 & 1.075$\pm.05$ \\ 
5982 & 24.25 & 0.365$\pm.07$ & 0.820$\pm.03$ & 0.500 & 1.100$\pm.04$ \\  
\hline
\end{tabular}
\end{flushleft}
\end{table}

\section{Conclusions}
New colour-radius relations have been derived for 36 E-type galaxies of the 
northern Local Supercluster, using UBVRI frames obtained with the 120cm telescope 
of the Observatoire de Haute-Provence. 
Four SA0, i.e. NGC3115, 3607, 4551 and 5866 were also observed. 

We aimed to take advantage of the large field of the camera  
to observe the galaxies at larger radii than hitherto feasible, and thus 
improve the accuracy of colour gradients. The availability of the series 
of aperture photometry in PP88 and PN94 for most of the sample was also 
considered an asset towards a more coherent system of colours. 
It appears indeed that {\em the colour calibrations are improved here 
compared to previous work}, if this can be judged from the quality of 
correlations between 
zero point colours in various surveys (see Sect. 4.1) 
 
Two steps in the reduction procedure were thought significant in improving 
the quality of colour profiles: the first was the adjustment of the FWHM 
of the PSFs in a given colour set of 5 frames to the best of the five. 
This allowed us to get significant colours much closer to the galaxy center 
than otherwise feasible. The second was a careful ''mapping" of the background 
of each frame, in order to lessen the background fluctuations remaining after 
the usual flat-fielding procedures. Both these precautions proved successful,  
and, as a result, the radial range of satisfactory colour measurements was 
greatly enlarged. Near the galaxy center, it proved feasible to obtain ''central 
colours'', i.e. colours integrated in the circle $r=3\arcsec$, in fair agreement 
with high resolution data (see Sect. 3.2.7.2 and Table 10). 

On the other hand, colours could be
obtained at much lower surface brightness (or larger radii) than in previous work. 
Our colour data extend to $23.2 <\mu_V <25.2$, with a median value near 
$\mu_V=24.5$ in all colours. According to the comparisons in Sect. 4.1, this is 
{\em 1.5 to 2 magnitudes deeper than in previous work}. ''External colours", 
refering to the level  $\mu_V=24.5$ whenever possible, are published for the 
first time (see Table 11), and may be useful to give some indications about  
stellar populations at the outskirts of E-galaxies. 

On the other hand, the ''red halo" effect of the camera was found to give 
enormous errors in V$-$I colours and gradients. These were corrected by a 
rigorous technique, and results in agreement with ''classical" data were 
obtained. Considering the V$-$I gradients, one is not happy however to introduce 
in their evaluation, corrections larger than the quantity to be measured!
Besides this specific problem of the red halo of thinned CCD, the far wings 
of the PSFs have been proven in a recent paper (RM01) to have non negligible 
effects in the gradients of other colours, and also to vary with the age of mirror 
coatings. It is not impossible that the U$-$B or U$-$V gradients given here 
are overestimated by 15-20\%, although they are in excellent statistical 
agreement with the well-known work of Peletier et al. (1990). 

Various colour gradients against $\log r$ for a given object are well correlated, 
generally better than in previous work (see statistics in Sect. 4.1), which is 
interpreted as due to smaller measuring errors, notably in U$-$B. 
These improvements in accuracy did not bring out any obvious correlation 
between gradients and other galaxy properties. A few galaxies have exceptionally 
steep colour gradients (nearly at $2\sigma$) without sharing other properties.   

Colour-colour relations can be built from the present data for several locations in
galaxies, such as near center, various fractions of the effective radius $r_e$, 
or the ''outermost" measured range around $\mu_V=24$. All these diagrams overlap 
to form a single stripe with moderate scatter (except for one rather obvious 
calibration error?). These might prove useful to test theories of old stellar 
populations and of their host galaxies. Colour-colour diagrams based upon 
integrated colours have already been used for this purpose (Worthey, 1994).  

The U$-$B or U$-$V colours correlate very well with the $Mg_2$ index, both near the 
galaxy center and at the effective radius $r_e$. This seems to rule out any 
large influence of diffuse dust in the colours and colour gradients in E-galaxies. 
This was considered likely by Witt et al. (1992) and discussed by Wise and Silva 
(1996) with inconclusive results. 
Previous arguments against such an influence were presented in RM00: they 
were based upon the relative average values of the gradients in various 
colours, and are reinforced in the present work, since the mean gradients are 
nearly unchanged, and their errors lessened. 

It is well known that, for single-burst stellar populations, colours and line 
indices depend both on the metallicity and on the age of the system (Worthey 1994; 
Borges et al. 1995). However, E-galaxies are constituted by a population mix, having
age and metallicity distributions which reflect their star formation histories. 
Therefore
a colour-metallicity calibration requires the use of models able to provide
those distributions and, consequently, the integrated colours along the galaxy lifetime.
Such a calibration will be presented in Paper II.

\begin{acknowledgements}
TPI acknowledges a Fapesp pos-doc fellowship n$^{\circ}$ 97/13083-7.
\end{acknowledgements}

\end{document}